\documentclass[preprint,12pt]{elsarticle}

\usepackage{graphicx}
\usepackage{multirow}
\usepackage[frozencache=true,cachedir=minted-cache]{minted} 
\usepackage{amsfonts,amsthm}
\usepackage{mathtools}
\usepackage{listings}
\usepackage{amsmath}
\usepackage{algorithm} 
\usepackage{algpseudocode} 
\usepackage{gensymb}
\usepackage{caption}
\usepackage{subcaption}
\usepackage{bera}
\usepackage{listings}
\usepackage[table,xcdraw]{xcolor}

\colorlet{punct}{red!60!black}
\definecolor{background}{HTML}{EEEEEE}
\definecolor{delim}{RGB}{20,105,176}
\colorlet{numb}{magenta!60!black}

\lstdefinelanguage{json}{
    basicstyle=\tiny,
    numbers=none,
    breaklines=true,
    frame=lines,
    backgroundcolor=\color{background},
    literate=
     *{0}{{{\color{numb}0}}}{1}
      {1}{{{\color{numb}1}}}{1}
      {2}{{{\color{numb}2}}}{1}
      {3}{{{\color{numb}3}}}{1}
      {4}{{{\color{numb}4}}}{1}
      {5}{{{\color{numb}5}}}{1}
      {6}{{{\color{numb}6}}}{1}
      {7}{{{\color{numb}7}}}{1}
      {8}{{{\color{numb}8}}}{1}
      {9}{{{\color{numb}9}}}{1}
      {:}{{{\color{punct}{:}}}}{1}
      {,}{{{\color{punct}{,}}}}{1}
      {\{}{{{\color{delim}{\{}}}}{1}
      {\}}{{{\color{delim}{\}}}}}{1}
      {[}{{{\color{delim}{[}}}}{1}
      {]}{{{\color{delim}{]}}}}{1},
}

\DeclareMathOperator*{\argmax}{arg\,max}
\DeclareMathOperator*{\argmin}{arg\,min}

\usepackage{hyperref}
\hypersetup{hidelinks}

\journal{Transportation Research Part C}

\begin{document}

\begin{frontmatter}



\title{What is a typical signalized intersection in a city? A pipeline for intersection data imputation from OpenStreetMap}

\affiliation[inst1]{organization={Department of Civil and Environmental Engineering (CEE), Massachusetts Institute of Technology},
            addressline={77 Massachusetts Avenue}, 
            city={Cambridge},
            postcode={02139}, 
            state={MA},
            country={U.S.A.}}
            
\affiliation[inst2]{organization={Department of Electrical Engineering and Computer Science (EECS), Massachusetts Institute of Technology},
            addressline={77 Massachusetts Avenue}, 
            city={Cambridge},
            postcode={02139}, 
            state={MA},
            country={U.S.A.}}

\affiliation[inst3]{organization={Institute for Data, Systems, and Society (IDSS), Massachusetts Institute of Technology},
            addressline={77 Massachusetts Avenue}, 
            city={Cambridge},
            postcode={02139}, 
            state={MA},
            country={U.S.A.}}

\affiliation[inst4]{organization={Department of Computer Science, Eidgenössische Technische Hochschule (ETH) Zürich},
            addressline={ETH Zürich}, 
            city={Zürich},
            postcode={8092}, 
            country={Switzerland}}

\author[inst1,inst5]{Ao Qu}

\author[inst2]{Anirudh Valiveru}

\author[inst2]{Catherine Tang}

\author[inst2,inst3]{Vindula Jayawardana}

\author[inst4]{Baptiste Freydt}

\author[inst1,inst2,inst3]{Cathy Wu, Ph.D.}


\begin{abstract}
Signalized intersections, arguably the most complicated type of traffic scenario, are essential to urban mobility systems. With recent advancements in intelligent transportation technologies, signalized intersections have great prospects for making transportation greener, safer, and faster. Several studies have been conducted focusing on intersection-level control and optimization. However, arbitrarily structured signalized intersections that are often used do not represent the ground-truth distribution, and there is no standardized way that exists to extract information about real-world signalized intersections. As the largest open-source map in the world, OpenStreetMap (OSM) has been used by many transportation researchers for a variety of studies, including intersection-level research such as adaptive traffic signal control and eco-driving. However, the quality of OSM data has been a serious concern. \\
In this paper, we propose a pipeline for effectively extracting information about signalized intersections from OSM and constructing a comprehensive dataset. We thoroughly discuss challenges related to this task and we propose our solution for each challenge. We also use Salt Lake City as an example to demonstrate the performance of our methods. The pipeline has been published as an open-source Python library so everyone can freely download and use it to facilitate their research. Hopefully, this paper can serve as a starting point that inspires more efforts to build a standardized and systematic data pipeline for various types of transportation problems. 
\end{abstract}


\end{frontmatter}

\section{Introduction}
The evolution of the web has allowed people with diverse cultural and geographic backgrounds to contribute to the same project through the rise of crowdsourcing. One significant example is OpenStreetMap (OSM), an open-source worldwide collaborative world map where users can contribute to a publicly-accessible database of geographic information~\cite{4653466}. As of January 2022, the project counts about 8.3 million users, 1.75 million of whom have been involved with making billions of edits to the platform. With such an abundance of accessible data, OSM has facilitated many academic studies, business and government operations, and daily usage, while becoming the only national-level source of spatial data in some countries~\cite{maplesotho,research_openstreetmap_wiki,open_cities_africa}. With its robust network of traffic infrastructure data, OSM has attracted growing interest from the transportation community and has been used for a variety of studies including travel demand modeling, transportation planning, traffic simulation and so on~\cite{klinkhardt2021using,zielstra2012using,zilske2015openstreetmap}. However, due to the lack of centralized efforts for quality assurance, OSM has suffered from inconsistent data quality and incomplete data labeling, issues that compound for traffic related data because of the heterogeneity of their formats and complexity of their linkages and attributes. Moreover, the required data elements, formats, and granularity may vary a lot depending on specific transportation research problems. Therefore, it is urgent to develop consistent and systematic methods for building datasets based on OpenStreetMap for different research problems.

If successful, such methods could unlock the potential of OpenStreetMap for transportation research.

Signalized intersections are significant to many critical issues such as sustainability, safety, and travel efficiency. In Delhi, around 15\% of fuel consumption is wasted by idling at signalized intersections~\cite{kumar2013idling}. According to a report released by the Federal Highway Administration (FHWA), signalized intersections account for about 21\% of all crashes and 24\% of all traffic fatalities~\cite{rodegerdts_2004}. There has been a long record of analyses performed on characteristics related to road intersections to improve transportation planning~\cite{doi:10.1177/0361198120919400,CAO202133,feng2020some,wang2018analyzing}. Recently, with an increasing investment in emerging technologies in intelligent transportation systems, smart signalized intersections have become a focus of study for experts in both academia and industry. Many studies and experiments have been conducted to push forward strategies such as autonomous intersection management (AIM), eco-driving, and adaptive traffic signal control (ATST) ~\cite{https://doi.org/10.48550/arxiv.2204.12561,9565053,wang2018review,6094668}. However, such studies often use arbitrarily fabricated intersection designs, making it difficult to evaluate their transferability. Some studies adapt real road networks, mostly from OSM, but the lack of standardization in feature extraction methods leads to difficulties in comparing research results. A recent study shows that the performance of traffic signal control algorithms is largely affected by the characteristics of signalized intersection scenarios, generally making the actual performance worse than claimed in the original source~\cite{jayawardana2022}. Without careful data curation and standardized methods for extracting signalized intersections from OSM, identifying the most efficient approach in smart intersection applications becomes quite challenging.

In our study, we thoroughly discuss challenges and issues that practitioners should bear in mind when extracting signalized intersections from OSM. To tackle all these challenges, we propose a systematic pipeline for collecting, merging, and imputing OSM data to build a signalized intersection dataset. To facilitate more studies related to signalized intersections, we have open sourced our tool as a Python library under active development (\url{https://pypi.org/project/OSMint/}). Although our study focuses on signalized intersections, the methods and imputation strategies discussed can easily be extended to several other portions of a road network, such as links and un-signalized nodes. To our best knowledge, this is the first paper that identifies the need for systematic approaches to extract intersection data for transportation research.

By proposing techniques to improve open source data quality, we hope this work will inspire further efforts to tailor systematic extraction pipelines to specific applications while ultimately increasing confidence in research that utilizes OSM data.

In the related work section, we will give a review of previous studies that assess OSM's data quality and provide a brief summary of potential applications of our methods. In the preliminaries section, we will cover important background knowledge pertaining to technical concepts that we discuss. In the methods section, we illustrate challenges related to building signalized intersection datasets and propose a solution for each one. In the results section, we analyze results obtained from our pipeline and evaluate its performance via a case study of Salt Lake City's road network. In the limitations and future work section, we discuss the limitations of the current study and future directions.

\section{Related Work}\label{related_work}
\subsection{OpenStreetMap and Data Quality}
OSM includes important street connectivity data to other streets, as well as crucial street information such as road types, lane counts, and speed limits. All data is crowdsourced, making it a popular source of intersection data for urban planning and intelligent infrastructure researchers. While OSM is the most popular dataset to adopt Wikipedia's crowdsourcing approach to compiling and collecting geodata, there are several commercial alternatives that exist, including Google Maps, Bing Maps, and HERE. As a case study comparing the accuracy of OpenStreetMap with the closed-source Google and Bing Maps in Dublin, Ireland, Ciepluch et al. concluded that all three datasets had similar sources of inaccuracy that made it difficult to determine the most accurate source out of the three~\cite{cipeluch}. More specifically, while OSM exhibited inaccurate or missing data at high-density locations due to a difficulty of finding volunteers willing to deal with high traffic, they found that Bing and Google Maps are prone to   errors in automated feature detection, with the added drawback of outdated data on Bing. While the accuracies of the most widely-used sources of geodata are equally indeterminate, OSM allows anybody to access their data for free, spurring the development of open-source frameworks like OSMnx for researcher to manipulate and visualize queried geodata~\cite{boeing2017osmnx}. OSMnx in particular was utilized significantly in the development of our methods to work with the complex graph network of Salt Lake City's streets; however, we found that its functionalities are limited for extracting intersections from OpenStreetMaps data. 

OSM's crowdsourced nature results in significant limitations to data availability (most research done on OSM's accuracy, for example, has only been done in the US, UK, and Germany), but the tool has become an essential way for researchers to apply street-level data to problems ranging from urban planning to intelligent transportation systems research~\cite{6822226}. One of this paper's main goals is to attempt curb the problem of data inavailability by proposing reasonable imputation methods that provide OSM users with access to data without missing values.

\subsection{Data-Driven Studies on Road Intersections}
Hentschel and Wagner integrated OSM data with regard to road types, speed limits, and connectivity in order to inform localization, path planning, and navigation tasks for autonomous robots in outdoor environments~\cite{5625092}. Similarly, Ballardini et al. utilized OpenStreetMap data as a ground truth data source to develop a probabilistic model for intersection detection by onboard cameras~\cite{7989030}. The Ballardini paper details one use of OSM that our paper particularly contributes to, being the use of intersection geometries for intelligent transportation research. While they only utilized information about the intersection's geometry, other studies such as Carlino et. al used lane counts, one-ways, and turn lane data to create traffic simulations using SUMO or AORTA~\cite{carlino}. Road intersection data from OSM, however, has some limitations that make it difficult to create such simulations without simplifying the graph. For example, Rieck et. al details several methods for tagging complex signalized intersections that are often represented as a single node in OpenStreetMap, which is necessary in order to use multi-lane intersections in simulation software like SUMO~\cite{rieck2015}. This is an important consideration for researchers that hope to create models that reflect the real world, and one that we aim to address in our work.

Overall, it is clear that road intersection models have a crucial role in transportation research, including emerging areas such as connected and automated vehicles. OpenStreetMap provides researchers with an easily accessible way to gain real-world information for modeling and analysis. Our paper hopes to contribute a framework using OSM to enable a low-cost high-quality intersection dataset for any city, using Salt Lake City as a case study for our methods of graph simplification and data imputation.

\section{Preliminaries}\label{prelim} 
\subsection{Data structures in OpenStreetMap}\label{osm_prelim} Most of the data from OpenStreetMap falls into three categories: nodes, ways, and relations\cite{OpenStreetMap}. These three types of data work together to provide a conceptual model of the physical world (e.g. road network).
\subsubsection{Node} A node represents a point on the earth's surface. It is represented as a 2-dimensional tuple containing its latitude and longitude. For example, a traffic signal can be modeled as a node in OpenStreetMap.
\subsubsection{Way} A way is an ordered list of more than two nodes. For example, a road segment is usually modeled as a way.
\subsubsection{Relation} A relation describes an interaction between specified nodes and ways. For example, a turn restriction is modeled as a relation, stating that a certain turn cannot be performed from way A to way B via node C.
\subsection{Graph}
A graph in OSM consists of nodes that are connected via ways. Mathematically, a graph is defined as $G=(V, E)$ where $V$ is the set of vertices and $E$ is the set of edges. The degree of a node denotes the number of edges connected, and each edge is connected to two distinct nodes unless it is a self-loop where the edge connects a node to itself. If there is a bijection between vertices from two graphs such that all the edges are preserved, they are considered isomorphic.
\subsection{Definitions related to traffic systems}\label{definitions}
\subsubsection{Intersection} 
\begin{figure}[!ht]
  \centering
  \includegraphics[width=0.6\textwidth]{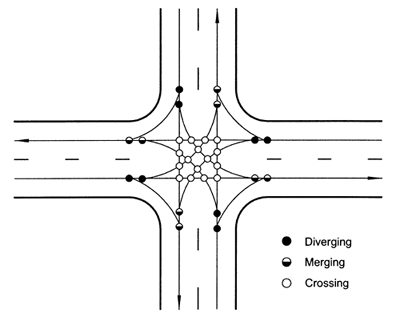}
  \caption{Intersection points in a signalized intersection with 4 approaches (US Department of Transportation)}
  \label{Fig:intersection-conflict}
\end{figure}

According to the U.S. Department of Transportation Federal Highway Administration, an intersection is defined as the area within which vehicles traveling on different highways that join at distinct angles may come into conflict~\cite{manual}. We define the overlap between two ways in OSM as a \textit{crossing}, to distinguish from an intersection. As figure \ref{Fig:intersection-conflict} shows, a typical intersection may include several crossings and involve many features.

\subsubsection{Signalized Intersection} A \emph{signalized intersection} is an intersection controlled by traffic signals. 
\subsubsection{Approach} Approaches are all lanes of traffic moving towards an intersection~\cite{manual}. Approaches and their associated attributes are the key to our dataset, since their properties are decisive to the traffic at an intersection.

\section{Methods}\label{methods}
\begin{figure}[!htt]
  \centering
  \includegraphics[width=0.9\textwidth]{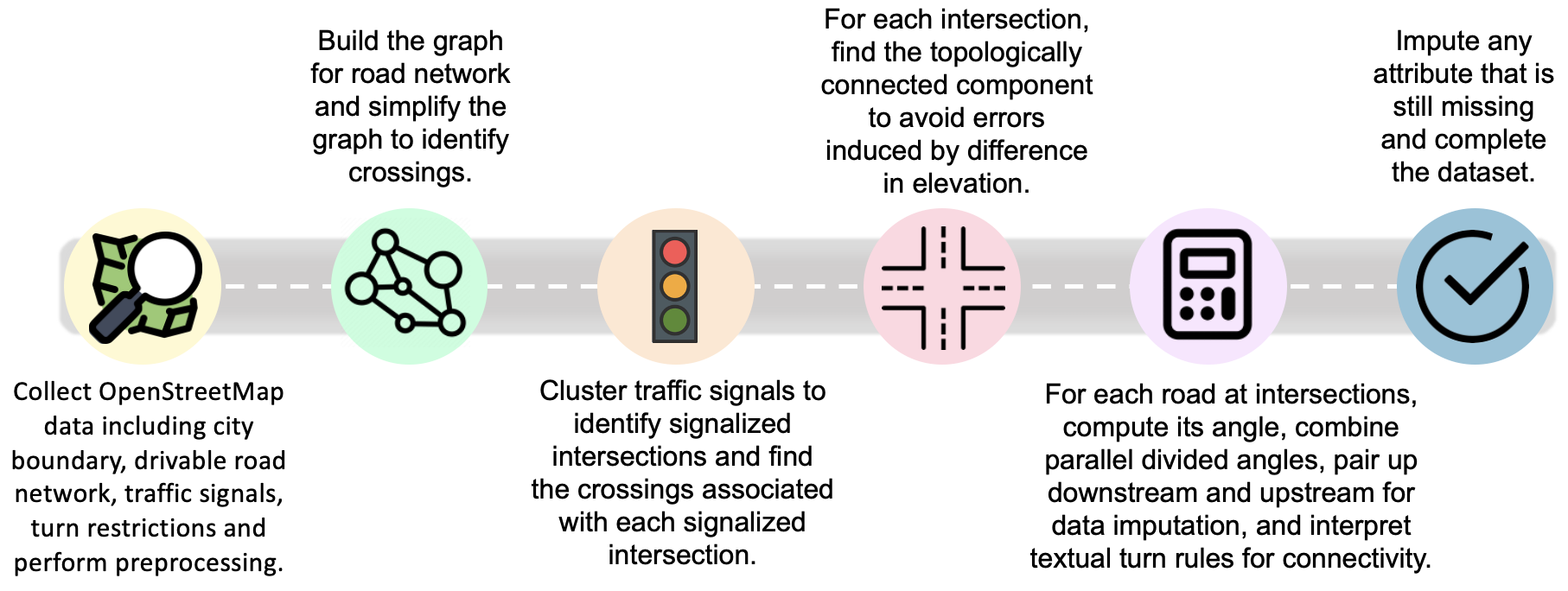}
  \caption{Pipeline for Extracting Signalized Intersections from OSM}\label{fig:pipeline}
\end{figure}

For multi-faceted applications like constructing signalized intersections, it is necessary to integrate differently structured data together and extract information from each data source for a comprehensive outcome. In this section, we initially propose our pipeline for collecting, processing, imputing, and integrating multiple types of data from OpenStreetMap. Then, we dive into details and discuss the challenges and issues associated with signalized intersection extraction. For each challenge, we propose a solution with algorithmic details and usage scenarios.\\
Figure \ref{fig:pipeline} shows a step-by-step summary of our pipeline. The first three steps focus on network-level operations, which decompose raw data into subsets associated with each signalized intersection identified in the third step. The next three steps are micro-scale operations that process data at intersection level and culminate in the well-organized signalized intersection dataset.

\hspace{-12mm}
\begin{minipage}[t]{0.35\linewidth}
\begin{lstlisting}[language=json]
{'type': 'way',
 'id': 1002466719,
 'nodes': [9251870521, 9251870508, 9251870520],
 'tags': {'highway': 'secondary',
          'lanes': '3',
          'lit': 'yes',
          'loc_name': '6th North',
          'maxspeed': '35 mph',
          'name': '600 North',
          'name:full': 'West 600 North',
          'name:prefix': 'W',
          'oneway': 'yes',
          'smoothness': 'excellent',
          'surface': 'asphalt',
          'turn:lanes': 'left||through;right'}}
\end{lstlisting}
\captionof{lstlisting}{A road segment example}
\label{listing:road}
\end{minipage}
\begin{minipage}[t]{0.31\linewidth}
\vspace{12mm}
\begin{lstlisting}[language=json]
 {'type': 'node',
 'id': 83516871,
 'lat': 40.7579328,
 'lon': -111.8299395,
 'tags': {'highway': 'traffic_signals', 
          'traffic_signals': 'signal'}
          }
\end{lstlisting}
\vspace{9mm}
\captionof{lstlisting}{A traffic signal example}
\label{listing:signal}
\end{minipage}
\begin{minipage}[t]{0.31\linewidth}
\vspace{5mm}
\begin{lstlisting}[language=json]
{'type': 'relation',
 'id': 1812091,
 'members': [
 {'type': 'way', 'ref': 10147035, 
 'role': 'to'},
 {'type': 'node', 'ref': 83592817, 
 'role': 'via'},
 {'type': 'way', 'ref': 346397931, 
 'role': 'from'}
 ],
 'tags': {'restriction': 'no_left_turn', 
          'type': 'restriction'}}
\end{lstlisting}
\vspace{3mm}
\captionof{lstlisting}{A turn restriction example}
\label{listing:restrictions}
\end{minipage}

\subsection{Data}
The three types of OSM data that we retrieved to construct the dataset are road segments, traffic signals, and turn restrictions. Examples of the raw data for each type are shown in Listings \ref{listing:road}, \ref{listing:signal}, and \ref{listing:restrictions}, respectively. Among them, road segment data is the most inconsistent in terms of attribute availability. In Table \ref{Table:summary_stats}, we display a summary of 16 significant road segment attributes in Salt Lake City, selected from 154 possible attributes defined in OSM. The second column shows the completeness of each attribute in Salt Lake City, while and the third column shows completeness for non-residential road segments. The fourth column shows an example of how each attribute may look for a specific road segment. Since a large portion of data is missing, some data imputation is necessary for a comprehensive intersection dataset.

\begin{table}[ht]
\resizebox{\textwidth}{!}{%
\begin{tabular}{|llll|}
\hline
\rowcolor[HTML]{EFEFEF} 
Attribute Name        & All road segments & W/O residential roads & Example                      \\ \hline
\rowcolor[HTML]{FFFFFF} 
type                  & 100.00\%          & 100.00\%              & way                          \\ \hline
\rowcolor[HTML]{FFFFFF} 
id                    & 100.00\%          & 100.00\%              & 107522082                    \\ \hline
\rowcolor[HTML]{FFFFFF} 
nodes                 & 100.00\%          & 100.00\%              & {[}4323899036, 9862965629{]} \\ \hline
\rowcolor[HTML]{FFFFFF} 
highway               & 100.00\%          & 100.00\%              & secondary                    \\ \hline
\rowcolor[HTML]{FFFFFF} 
name                  & 85.60\%           & 78.68\%               & South 5600 West              \\ \hline
\rowcolor[HTML]{FFFFFF} 
lanes                 & 54.00\%           & 82.83\%               & 5                            \\ \hline
\rowcolor[HTML]{FFFFFF} 
lanes:backward        & 7.07\%            & 11.96\%               & 2                            \\ \hline
\rowcolor[HTML]{FFFFFF} 
lanes:both\_ways      & 3.05\%            & 5.29\%                & 1                            \\ \hline
\rowcolor[HTML]{FFFFFF} 
lanes:forward         & 7.07\%            & 11.93\%               & None                         \\ \hline
\rowcolor[HTML]{FFFFFF} 
oneway                & 40.41\%           & 65.28\%               & no                           \\ \hline
\rowcolor[HTML]{FFFFFF} 
maxspeed              & 43.18\%           & 63.87\%               & 45 mph                       \\ \hline
\rowcolor[HTML]{FFFFFF} 
incline              & 1.03\%           & 1.05\%               & 15\%                       \\ \hline
\rowcolor[HTML]{FFFFFF} 
turn:lanes            & 5.91\%            & 10.25\%               & None                         \\ \hline
\rowcolor[HTML]{FFFFFF} 
turn:lanes:both\_ways & 3.05\%            & 5.29\%                & left                         \\ \hline
\rowcolor[HTML]{FFFFFF} 
turn:lanes:backward   & 1.87\%            & 3.09\%                & through|right                \\ \hline
\rowcolor[HTML]{FFFFFF} 
turn:lanes:forward    & 2.21\%            & 3.58\%                & left;through|None            \\ \hline
\end{tabular}%
}
\caption{Summary of Road Segments Attributes}
\label{Table:summary_stats}
\end{table}
\subsection{Identification of signalized intersections} \label{identification}
According to the definition of signalized intersections given in Definitions, we model a signalized intersection as a graph such that the area in which conflicts may occur is consolidated as a single node, while approaches are modeled as edges connected to this node. This structure is a tree, with the intersection node acting as the root and each connected approach edge representing a leaf node. Our goal is to extract all possible tree structures with along with associated important edge attributes (e.g. speed limit, lane count, turns, etc).

The challenges for identifying signalized intersections using OSM data are twofold. First, traffic signal data can only be retrieved separately from road network data, so matching is required to find the intersections that have traffic signals installed. Also, one intersection may contain several traffic signals (e.g., separate signals for individual approaches) so it is necessary to cluster them into groups. Second, identifying intersections itself is challenging. In OSM, incoming and outgoing approaches can be inconsistently encoded as either the same way or two separate ways, even if they connect to an intersection at the same angle. Therefore, naively identifying the overlap between ways leads to an overestimation of the actual number of intersections. As pointed out in previous studies, the joining of two \textit{divided highways} will sometimes be encoded as four crossings in OSM. A common approach is to merge neighboring crossings within a certain threshold. However, this approach can be problematic as some intersections connecting to motorways span large distances and a threshold that accommodates such intersections would erroneously lump together smaller intersections in urban areas where intersections are more densely located.

To tackle these two challenges, we first perform clustering on traffic signals. The problem with clustering crossings is that distances between intersections can be inconsistent. However, more consistent designs for signalized intersections are maintained. For instance, according to \cite{njroadway}, spacing between signalized intersections is suggested to be at least 1200 feet. Despite the potential for discrepancies in this suggestion, we should be much more confident about setting a threshold for clustering traffic signals. The result is a mapping $f: S \rightarrow M$ such that $f(S_i)=f(S_j)$ if $\texttt{dist}(S_i, S_j) < \delta$ where $S$ denotes traffic signals, $M$ denotes clusters, and $\texttt{dist}$ is the distance metric (i.e. Euclidean) with the threshold denoted by $\delta$.

\begin{figure}[!ht]
  \centering
  \begin{subfigure}{.3\textwidth}
  \centering
  \includegraphics[width=1\textwidth]{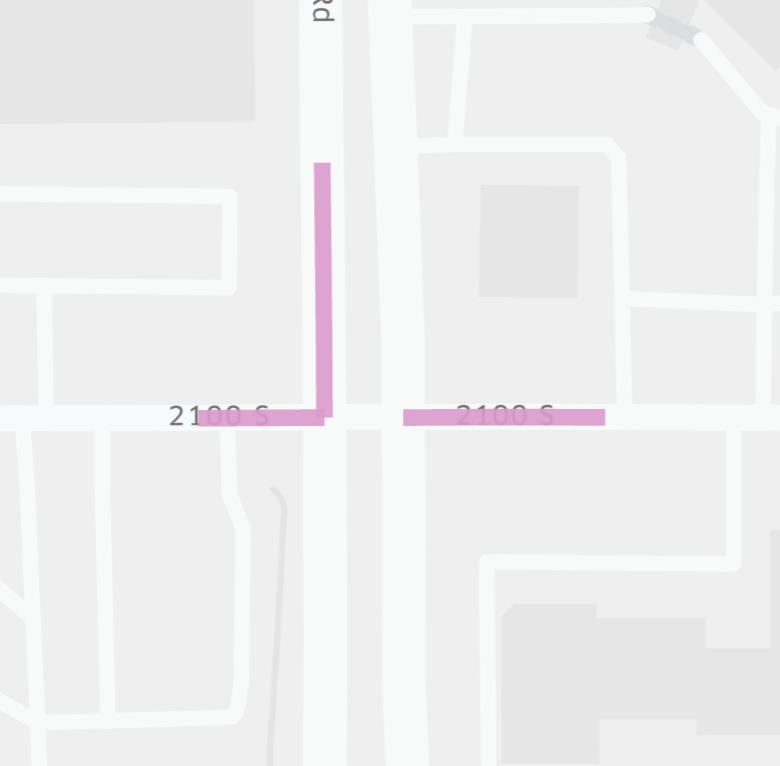}
  \caption{Road segments connected to traffic signals}\label{Fig:connection:original}
  \end{subfigure}
  \begin{subfigure}{.3\textwidth}
  \centering
  \includegraphics[width=1\textwidth]{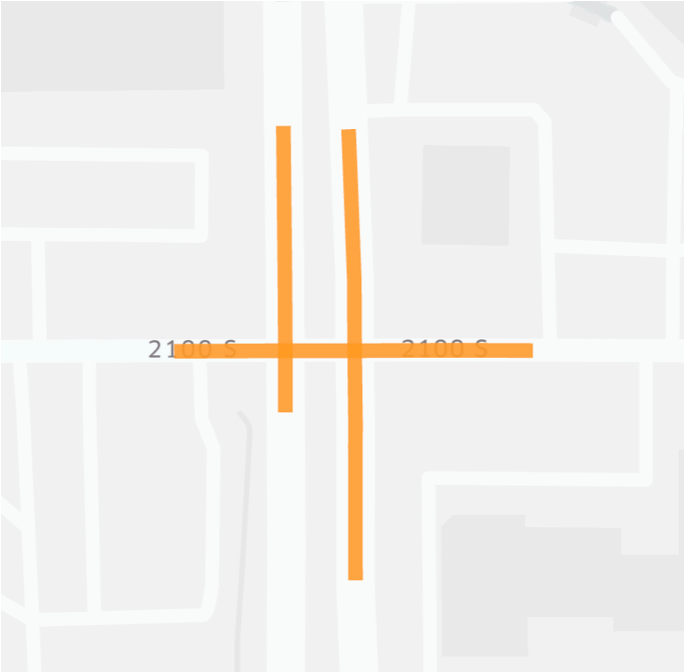}
  \caption{Road segments found using our approach}\label{Fig:connection:new}
  \end{subfigure}
  \begin{subfigure}{.34\textwidth}
  \centering
  \includegraphics[width=1\textwidth]{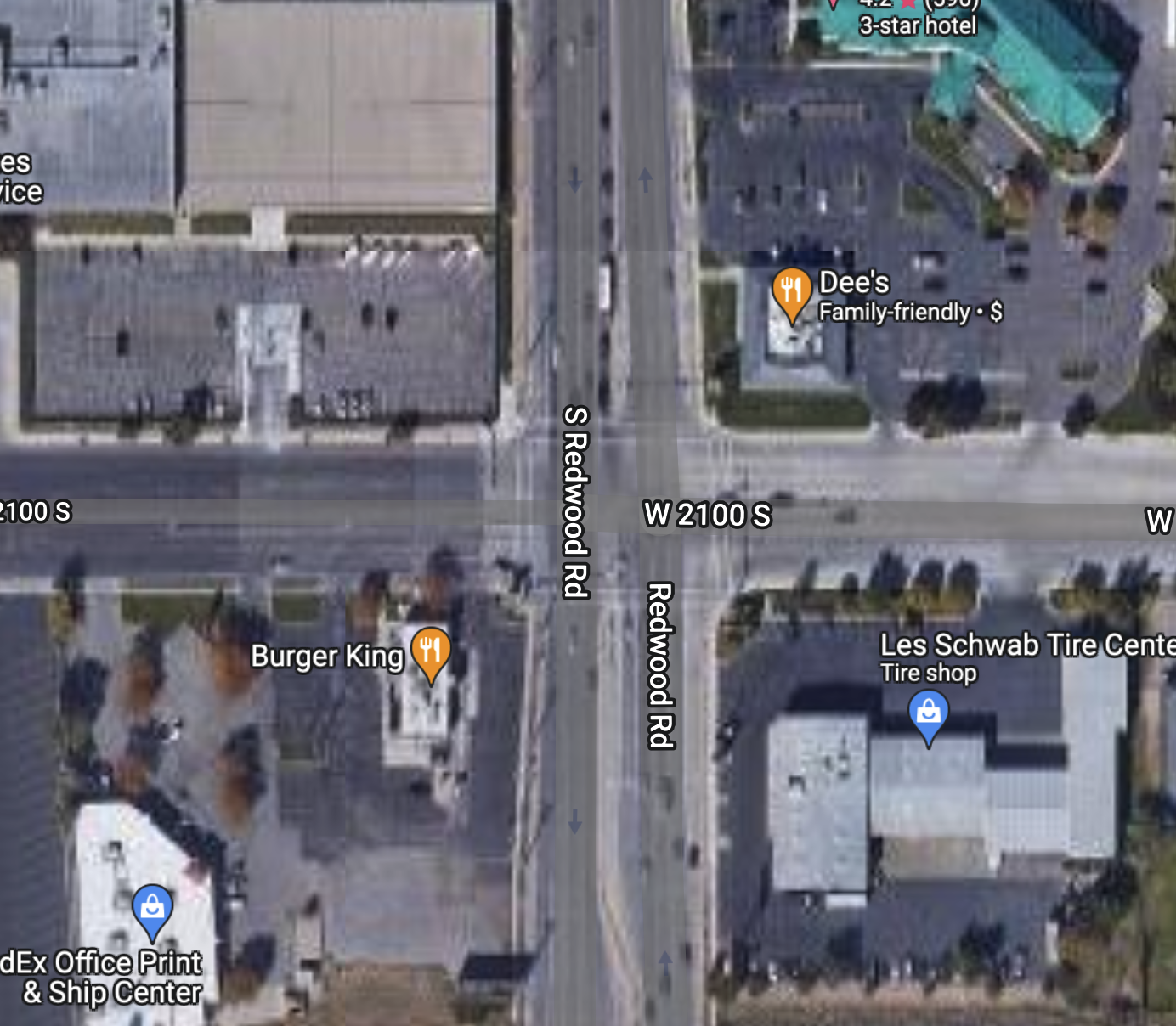}
  \caption{Satellite view of the signalized intersection}\label{Fig:connection:satellite}
  \end{subfigure}
  \caption{An example of identifying a signalized intersection with its approaches.}
  \label{Fig:identification}
\end{figure}

As illustrated in Figure \ref{Fig:identification}, some roads that run through a signalized intersection do not actually pass through traffic signals. Therefore, we must explicitly identify all the crossings near a traffic signal cluster. Now, we can use a much smaller distance threshold for matching crossings to traffic signal clusters, since a larger signalized intersection $i$ tends to have more traffic signals, or a larger space spanned by $|f^{-1}(M_i)|$ with $M_i$ denoting the cluster for $i$, which automatically leads to a larger range for identifying crossings. Let us denote a way $W_i$ as an ordered sequence $(N_1^i, N_2^i, \dots, N_n^i)$.  To find all the crossings, we can simply construct a graph $G=(V, E)$ where $V=\{N_i^j: \exists W_j, N_i^j\in W_j\}$ and $E=\{(N_i^j, N_{i+1}^{j}): \exists W_j, N_i^j, N_{i+1}^j\in W_j\}$. Then, the set of crossings, $C$, will be all nodes $N_i^j$ in $G$ such that $\texttt{degree}(N_i^j) > 2$. Then, we can use a much smaller threshold $\epsilon$ for mapping crossings to traffic signal groups. In particular, a mapping $g: C \rightarrow G$ is defined as follows:
$(\exists S_m, S_n\in S, C_i, C_j\in C), \texttt{dist}(C_i, S_m)<\epsilon, \texttt{dist}(C_j, S_n)<\epsilon, f(S_m)=f(S_n)=M_k\Rightarrow g(C_i)=g(C_j)=M_k$
In practice, this search can be efficiently done using tree-based methods such as KD-tree and R-tree. To retrieve all approaches for a signalized intersection $i$ we identified, we can simply perform the following three steps. First, we find all the crossings occurred around $i$, which can be denoted as $g^{-1}(M_i)$ and we call it the set of intersection nodes. Second, we use the OSM ways that run through nodes in $g^{-1}(M_i)$, denoted as $\{W_k: \exists N_j^k\in W_k, N_j^k\in g^{-1}(M_i)\}$, to construct a graph $G_i$ similarly as we did for all ways. Third, in order to address errors caused by differences in elevation (e.g. crossings on Overpass above $i$ should not be considered as a part of $i$), we need to find the largest connected component of the associated graph $G_i$. The roads contained by this component will be the approaches for $i$. \\
This approach helps us more accurately identify roads that make up a signalized intersection and effectively address concerns related to inflexible thresholds and false connectivity. For the rest of this section, we will discuss micro-level operations to organize and complete details for each intersection.

\subsection{Recovery of intersection geometry}
\begin{figure}[!ht]
  \centering
  \begin{subfigure}{.31\textwidth}
  \centering
  \includegraphics[width=1\textwidth]{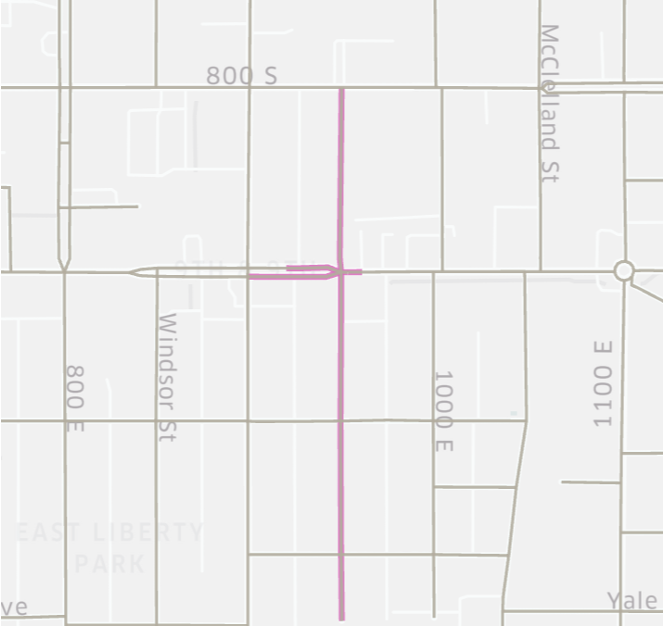}
  \caption{Original geometry of road segments from OSM}\label{Fig:connection:original1}
  \end{subfigure}
  \begin{subfigure}{.31\textwidth}
  \centering
  \vspace{5mm}
  \includegraphics[width=1\textwidth]{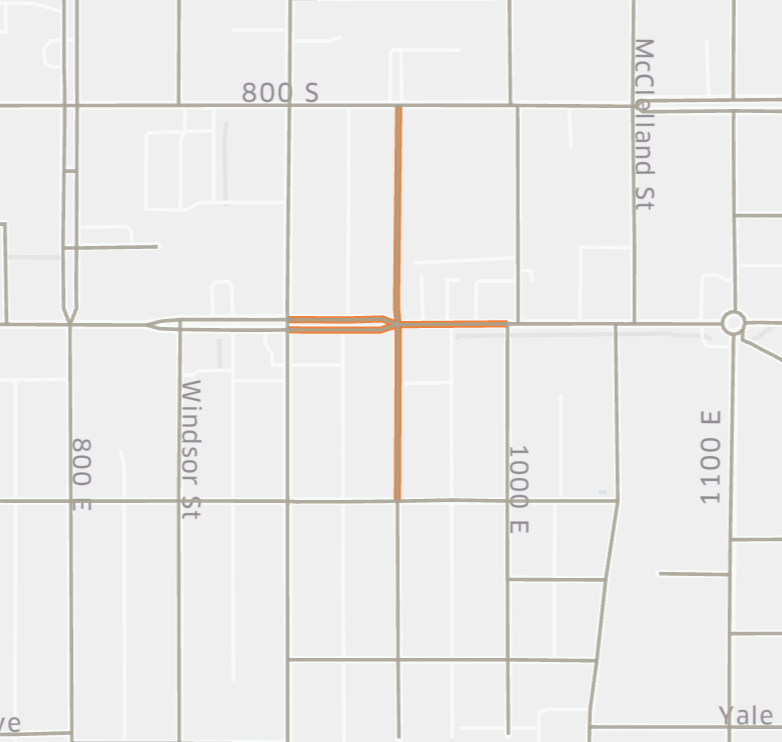}
  \caption{Recovered geometry with approaches ending at the closest intersections}\label{Fig:connection:new1}
  \end{subfigure}
  \begin{subfigure}{.3\textwidth}
  \centering
  \includegraphics[width=1\textwidth]{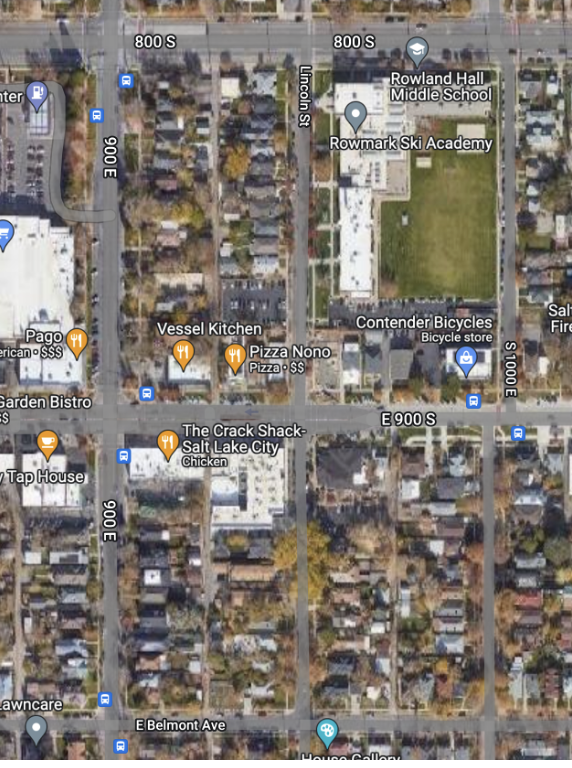}
  \caption{Satellite view of the signalized intersection}\label{Fig:connection:satellite1}
  \end{subfigure}
  \caption{An example of recovering geometry for approaches}
  \label{Fig:recovery}
\end{figure}

A challenge with how OSM defines ways is the lack of a consistent rule about how road segments are split. Since anyone can contribute to OSM, a split may occur simply when a partially complete road segment is edited by another user. As illustrated in Figure \ref{Fig:recovery}, naively using the geometries of road segments returned from OSM does not accurately reflect the actual length and shape of each approach. Since an intersection controls the traffic along an approach up until the next intersection, developing a method to retrieve complete geometries for each approach is integral to our goals. To achieve this, we start with graph $G$ for the complete road network, built following procedures described in \ref{identification}. Then, to get the actual length and shape for each approach, we initially simplify the graph by merging all nodes with degree two. An algorithm for simplifying the graph is described in Algorithm \ref{graphsimp}. In this algorithm, we keep a map to track nodes that have been taken from each newly formed edge. Two keys are stored for each edge in order to ensure the order of stored sequences. Having this map filled, once we identify that a road segment $W_i = (N_1^i, \dots, N_2^i)$ is an approach for the signalized intersection $j$, we can find $(N_k^i, N_{k+1}^i)$ such that exactly one of them is an element in $g^{-1}(M_i)$. Although many road segments end at intersections, some just go through intersections so that it is possible to have one road segment cover both upstream and downstream, therefore resulting in two such $(N_k^i, N_{k+1}^i)$ pairs. To address this issue, we intentionally split such roads during the data processing stage. After such $k$ is found for $W_i$, we can find edge $(u, v)\in E(G)$ such that \texttt{nodeDict}[(u, v)] contains $N_k^i$ and $N_{k+1}^i$ and \texttt{nodeDict}[(u, v)] shall be used to recover the actual shape and length of this approach. In particular, if we denote \texttt{nodeDict}[(u, v)] as $(x_1, x_2, \dots, x_n)$ where each $x_i$ is a node associated with its latitude (Lat) and longitude (Lon), the actual shape and length of the approach can be recovered as follows,
$$shape = [(x_1.Lat, x_1.Lon), \dots, (x_n.Lat, x_n.Lon)]$$
$$length = \sum_{i=1}^{n-1} \texttt{dist}(x_i, x_{i+1})$$
Again, \texttt{dist} is the distance metric. 

\begin{algorithm}
	\caption{Road Network Simplification Algorithm} 
	\label{graphsimp}
	\begin{algorithmic}[1]
		\State \texttt{neighbor} returns all the neighbors of a given node. 
        \State \texttt{remove} removes a node from G along with all edges connected to that node.
        \State \texttt{concat} concatenates two sequences of nodes.
        \State \texttt{reverse} returns a node sequence in reverse order.
        \State Initialize \texttt{nodeDict}, a map used to store nodes removed for new edges. 
		\For{(u, v) in E(G)}
            \State \texttt{nodeDict}[(u, v)] = (u, v)
            \State \texttt{nodeDict}[(v, u)] = (v, u)
        \EndFor
		\State biNodes = $\{N\in V(G): degree(N)==2, |\texttt{neighbor}(N)|>1\}$
    
        \While {biNodes is non-empty} 
            \For {node in biNodes}
                \State neighbor1, neighbor2 = \texttt{neighbor}(node)
                \State add (neighbor1, neighbor2) to E(G)
                \State remove(node)
                \State \texttt{nodeDict}[(neighbor1, neighbor2)] = \texttt{concat}(\texttt{nodeDict}.pop((neighbor1, node)), \texttt{nodeDict}.pop((node, neighbor2)))
                \State \texttt{nodeDict}[(neighbor2, neighbor1)] = \texttt{reverse}(\texttt{nodeDict}[(neighbor1, neighbor2)])
            \EndFor
    \EndWhile
    \State return G, \texttt{nodeDict}
	\end{algorithmic} 
\end{algorithm}

\subsection{Approach angle}

\begin{figure}[!ht]
  \centering
  \begin{subfigure}{.45\textwidth}
  \centering
  \vspace{10mm}
  \includegraphics[width=1\textwidth]{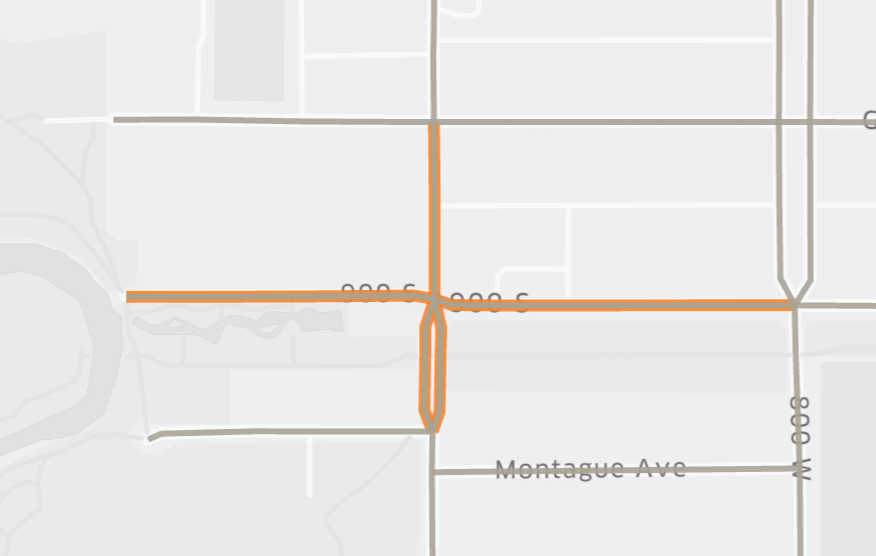}
  \vspace{10mm}
  \caption{This one requires "far-sighted" calculation to identify the two antiparallel road segments}\label{Fig:curve1}
  \end{subfigure}
  \begin{subfigure}{.35\textwidth}
  \centering
  \includegraphics[width=1\textwidth]{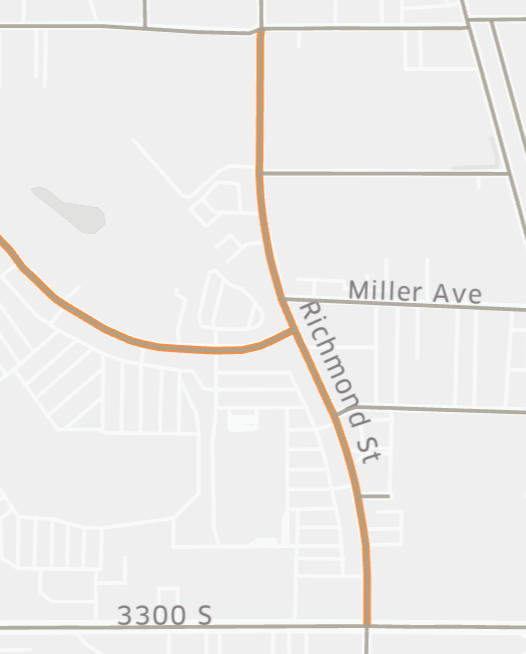}
  \caption{This example requires "short-sighted" calculation to interpret turn information correctly}\label{Fig:curve2}
  \end{subfigure}
  \caption{Two examples demonstrating why we need to consider two types of angles}
  \label{Fig:angles}
\end{figure}

Angles of each approach are crucial features to a signalized intersection. Traffic patterns can differ significantly for intersections with different approach angles even if their graph structures are isomorphic. For instance, a T-shaped intersection is functionally very different from a Y-shaped intersection, despite both having three approaches. When handling OSM data, angle information is even more important to identify two-way roads that are classified as two one-ways and pair downstream edges with upstream edges of the same road when imputing data. As mentioned in the previous sections, we have been able to find road segments that are connected to a signalized intersection $i$, perform necessary data processing, and find $(N_k^j, N_{k+1}^j)$ for each road segment $W_j$ such that exactly one of them is an intersection node. Then, to compute the angle, we first project all coordinates in latitude and longitude to the Universal Transverse Mercator (UTM) coordinate system for accuracy and ease of computation. Without loss of generality, we assume $N_{k+1}^j$ to be the intersection node and the angle is computed as 
$$\theta_j = arctan(\frac{N_k^j.y - N_{k+1}^j.y}{N_k^j.x - N_{k+1}^j.x})$$
where $x,y$ are UTM coordinates. Since $arctan$ ranges from $-90^{\circ}$ to $90^{\circ}$, we let $\theta_j = 180 - \theta_j$ if $N_k^j.x < N_{k+1}^j.x$ and then add $360$ to $\theta_j$ if it is negative to ensure that $\theta_j$ has a range of $[0^{\circ}, 360^{\circ})$. As illustrated in Figure \ref{Fig:angles}, for some signalized intersections, roads do not straightly run into the intersection and our "short-sighted" method for calculating theta may not generate accurate results. Therefore, we also consider a "far-sighted" calculation for theta that finds the angle of the next crossing along the non-intersection node to the intersection node. For our applications, we use both angle measures to maximize accuracy.

\subsubsection{One-way or Not?}
In OSM, many roads that could have been represented as a single two-way are drawn as two one-ways, even if they are not physically divided. Since the physical barrier disappears at the intersection area regardless, we combine such pairs of one-ways into two-ways to get more a accurate representation of each intersection. Without this step, there is no way to distinguish between a three-way intersection with one divided edge and a typical four-way intersection. To identify such pairs, we compute the difference between approach angles and group the edges that have a difference less than a certain threshold. In particular, we combine two road segments $W_m, W_n$ that are connected to a signalized intersection if 
$$\min{(\angle(\theta_{short}(W_m), \theta_{short}(W_n)), \angle(\theta_{long}(W_m), \theta_{long}(W_n)))}<\epsilon_{one-way}$$
where $\min$ takes the minimum of two, $\angle$ calculates the acute angle between two angles, $\theta_{short}, \theta_{long}$ denote two types of approach angles as introduced above, and $\epsilon_{one-way}$ denotes the threshold. After matching, the generated road segments or pairs of road segments should represent the actual approaches of an signalized intersection. 
\subsubsection{Road lane information}
\begin{figure}[!ht]
  \centering
  \begin{subfigure}{.5\textwidth}
  \centering
  \includegraphics[width=1\textwidth]{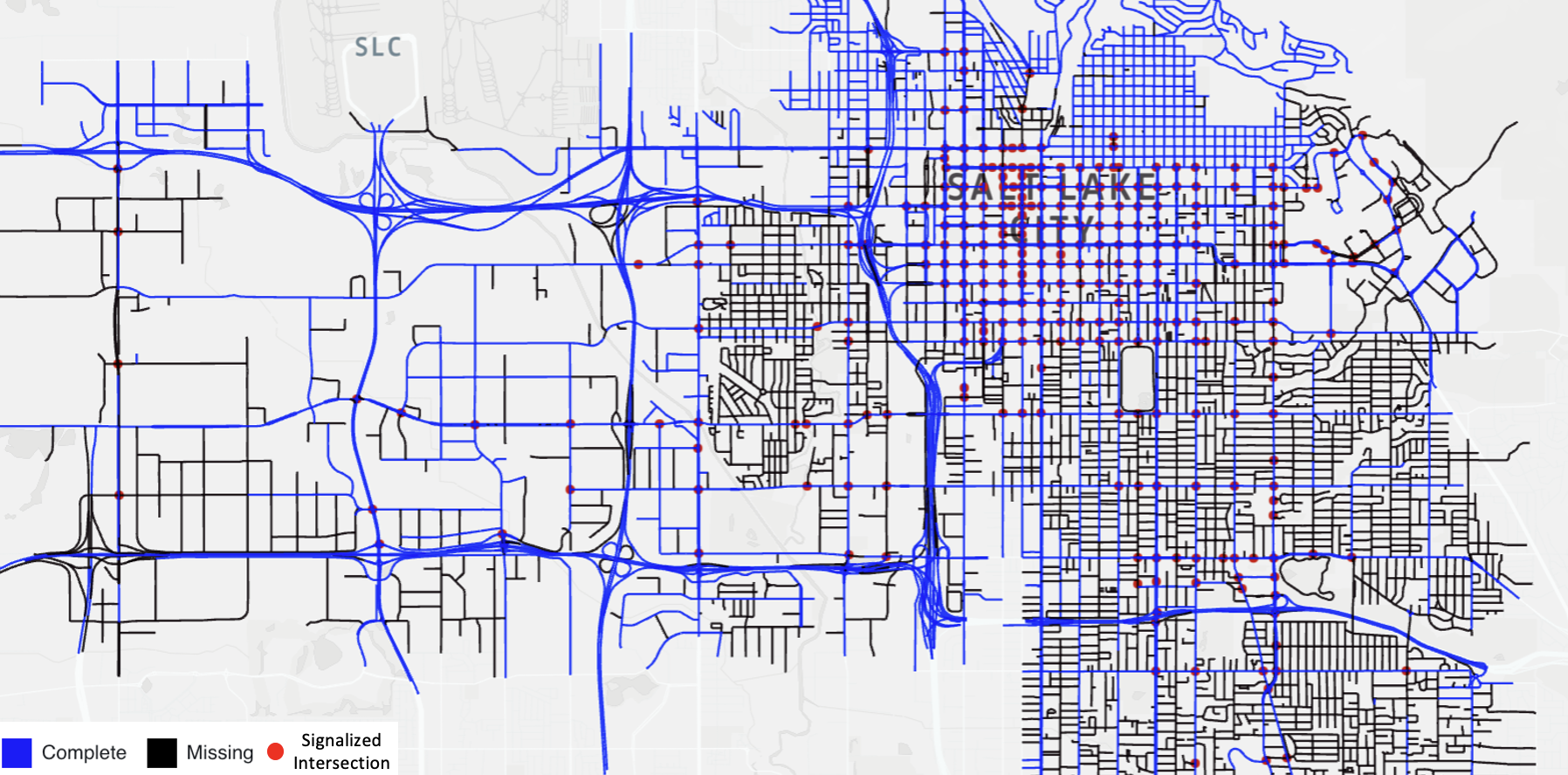}
  \caption{Availability for Lane Count in Salt Lake City}\label{fig:lane_count}
  \end{subfigure}
  \begin{subfigure}{.5\textwidth}
  \centering
  \includegraphics[width=1\textwidth]{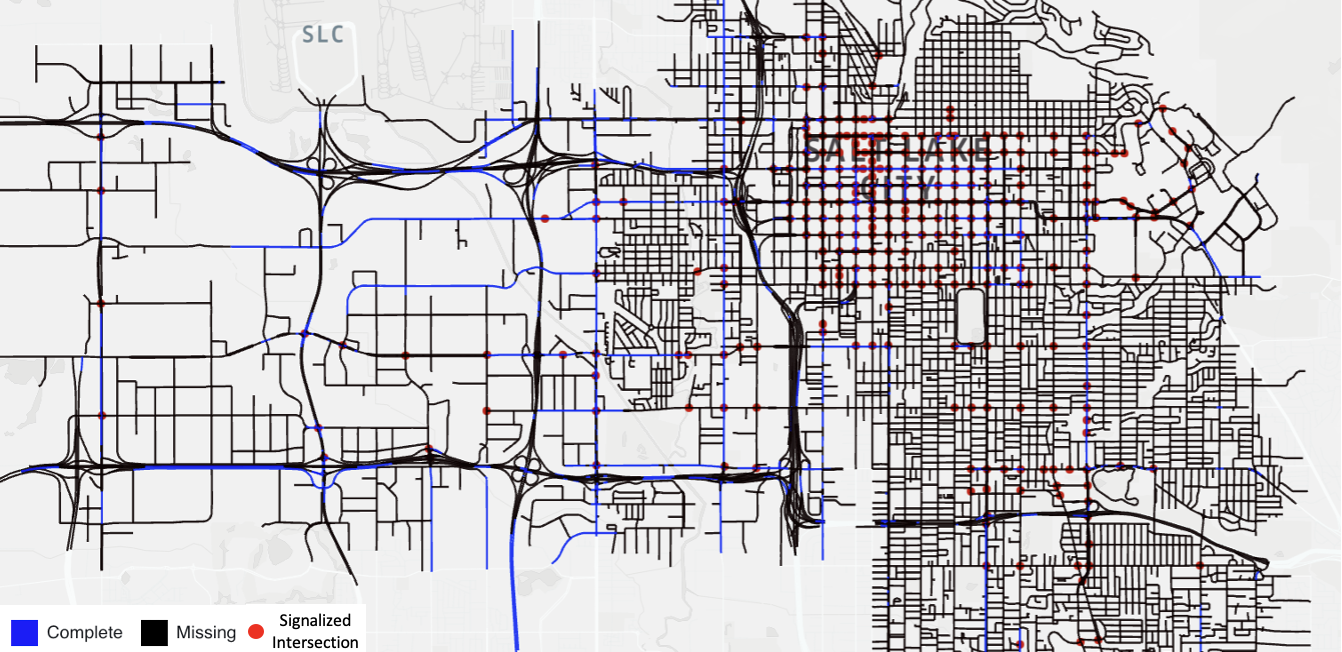}
  \caption{Availability for Turn Data in Salt Lake City}\label{fig:turn}
  \end{subfigure}
  \begin{subfigure}{.45\textwidth}
  \centering
  \includegraphics[width=1\textwidth]{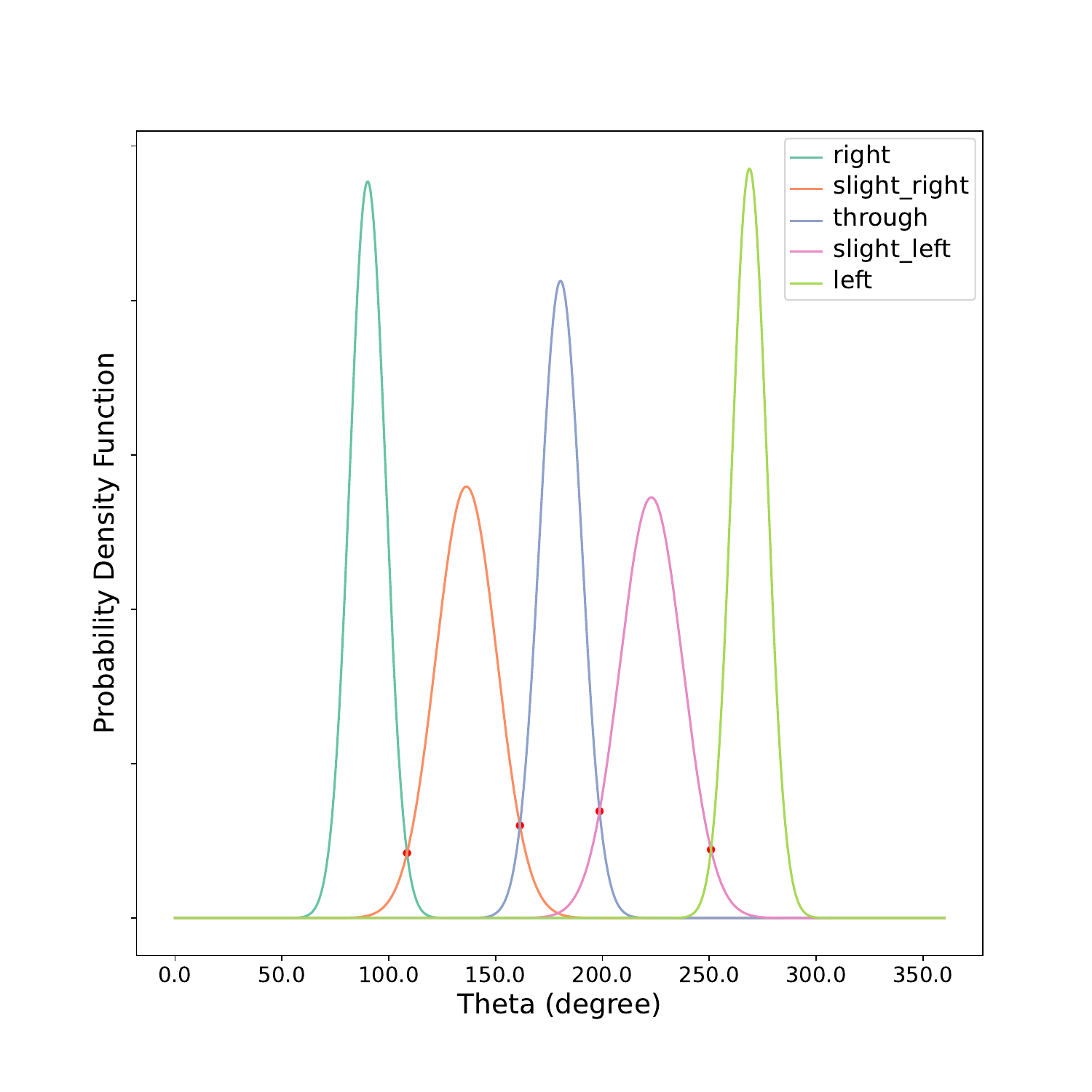}
  \caption{Angle Distributions for Different Turn Types in Salt Lake City}\label{fig:turn_distribution}
  \end{subfigure}
\end{figure}

As shown in Figure \ref{fig:lane_count},  54.1\% of roads in Salt Lake City do not have lane count information. Such incompleteness is very common for OSM data, motivating the need to find an effective way to impute lane counts. In most cases, downstream and upstream edges of a road going through an intersection should have the same lane count. Therefore, once we can pair a downstream road segment with its corresponding upstream road segment, we can leverage the information from both sides for lane count imputation. Similarly to how we combine anti-parallel roads, two road segments $W_m, W_n$ are determined to be an upstream-downstream pair if 
$$\max{(\angle(\theta_{short}(W_m), \theta_{short}(W_n)), \angle(\theta_{long}(W_m), \theta_{long}(W_n)))}>\epsilon_{updown}$$
where $\epsilon_{updown}$ denotes the threshold to identify roads that may form a through traffic as opposed to a turn. 

It is also possible that a two-way road becomes a one-way after the intersection. For such scenarios, we will only match lane count information for the part of two-way with the same direction as the one-way. If any approach still misses lane count, we perform an imputation merely based on its road type and fill in the blank with the most frequently occurring lane count for its type.

\subsection{Turning Lanes}
Allowing drivers to navigate through urban areas with great flexibility, turn lanes are essential to signalized intersections. In OSM, there are two ways to extract available turn information--the textual turn lanes specified with each road segment, or turn restrictions that need to be retrieved separately. Our goal is to obtain complete information about connectivity between all lanes. However, there are challenges associated with interpreting each of these two data sources. 
\subsubsection{Textual Turns}
Processing textual turn data from OSM is challenging because terms such as "left", "slight left", and "through" are ambiguous for some irregularly shaped intersections. To process this information in a systematic way, we first draw a random sample for each type of turn and fit the turn angles into a normal distribution. Figure \ref{fig:turn_distribution} shows an example of the distributions we obtained for Salt Lake City. Using this distribution, we convert a textual turn data with type $A$ from approach $W_i$ into connectivity information by finding the approach $W_j$ such that
$$W_j = \argmax_{W\in \text{all approaches}} f(\angle{(\theta(W_i), \theta(W))}|(\mu_{A}, \sigma_{A}))$$
where $\mu_A, \sigma_A$ are mean and standard deviation for the normal distribution fitted to approach angles of turn type $A$ and $f$ is the probability density function for normal distribution which takes the form $f(x|\mu, \sigma) = \frac{1}{\sqrt{2\pi}\sigma}e^{-\frac{1}{2}(\frac{x-\mu}{\sigma})^2}$. However, turn data is inherently missing for many road segments in OSM as shown in Figure \ref{fig:turn}. For such approaches, we simply use the default setting, which lets the leftmost lane turn left, the rightmost lane turn right, and all other lanes must go straight.
\subsubsection{Turn Restrictions}
Although they are stored separately from road segments in OSM, turn restrictions are also helpful for us to infer connectivity. As mentioned in section \ref{osm_prelim}, turn restrictions are represented as relations in OpenStreetMap. For instance, a no-left-turn restriction from way $W_i$ to way $W_j$ via node $N_k$ means that a left turn from $W_i$ to $W_j$ via $N_k$ is not allowed. However, most restrictions are manually added by OSM users who lack formal training and have created new ways and nodes that are not present in road segment data. Therefore, we must perform a matching to integrate turn restrictions to the rest of the dataset. Since turn restriction data is not linked with road segment data for each signalized intersection, we first need to find turn restrictions that have its \texttt{Via} part topologically close to the signalized intersection. Then, for \texttt{From} and \texttt{To} parts, we match them with approaches that have been identified for the signalized intersection by computing their angular distance regarding the intersection. In particular, for a \texttt{From} way $W_{from}$, we can find its match $\hat{W}_{from}$ in existing intersection approaches by the following equation, 
$$\hat{W}_{from} = \argmin_{W\in \text{all approaches}} \angle(\theta_{r}(W_{from}), \theta_{short}(W))$$
where $\theta_{r}$ calculates the angle for $W_{from}$ regarding \texttt{Via} because $W_{from}$ may not even contain any intersection node so we use \texttt{Via} to represent the location of the turn restriction more accurately. A similar procedure is performed to match \texttt{To} way. After the matching is done, we simply remove any conflicting connectivity that has been established.

\subsection{Speed Limit Imputation}
One key feature of intersections that we want to encode in our intersection distributions is speed limits, which are crucial to understanding the dynamics of traffic at an intersection. One significant challenge we ran into when developing a method to compile speed limits in Salt Lake City was the scarcity of concrete speed limit data available on both OSM and other mapping APIs like Google Maps, HERE Maps, and TomTom. Because of this, we had to make several assumptions that would allow us to predict a reasonable speed limit for every street in Salt Lake City. The first assumption is that Open Street Maps' \emph{maxspeed} attribute is accurate to the ground truth of speed limits; the second assumption is that lane count and road type are both crucial in determining the speed limit of an unknown edge; the third assumption is that the default speed limit of all roads (that cannot be imputed by previous methods) is determined by Utah's \emph{prima facie} speed limit laws, which state that residential streets have a speed limit of 25 miles per hour and highways have a speed limit of 70 miles per hour. 

Taking these assumptions into account, the pipeline for speed limit imputation is as follows:

1. Impute residential streets (as determined by the \emph{highway} property) based on the state's transportation code. Impute motorways in a similar way.

2. For any edges that have a null speed limit after the first step, we impute the speed limit based on road name. If more than 90\% of non-null edges with a particular street name have the same speed limit, then we assume that all null edges along that road also correspond to said speed limit. This is essentially a "default" speed limit for any edges along that road.

3. We now look at all possible combinations of lane count and highway type. Since lane count is not available for every edge in the OSM network, this combination is only done on non-imputed lane counts for the sake of accuracy. For each combination, impute all null edges with the most frequent speed limit at the given combination. By the end of this step, all streets with lane counts should be imputed

4. The last step is to look at all highway types (without lane counts). On this step, we impute all null edges for a given highway type with the most frequent speed limit for that street type.

\begin{figure}[ht]
  \centering
  \begin{subfigure}{.55\textwidth}
  \centering
  \includegraphics[width=1\textwidth]{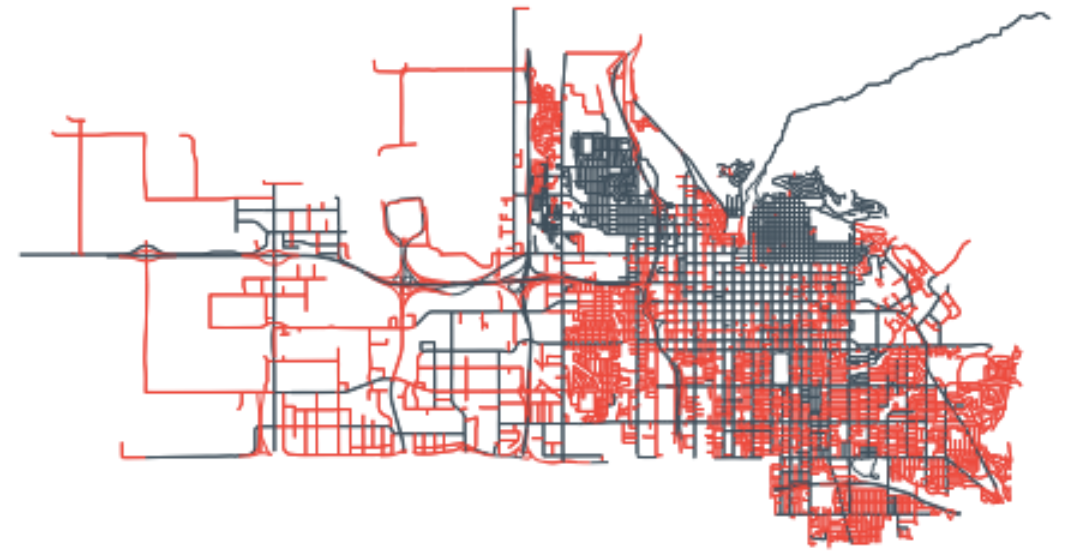}
  \caption{Availability for Speed Limit in Salt Lake City (Red is null, Gray is available)}\label{fig:maxspeed}
  \end{subfigure}
  \begin{subfigure}{.45\textwidth}
  \centering
  \includegraphics[width=0.9\textwidth]{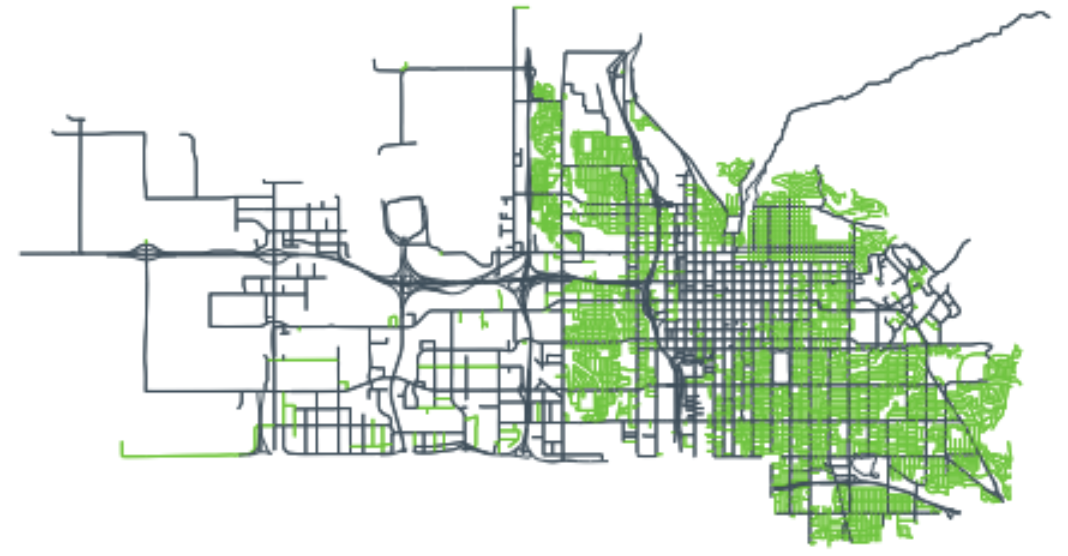}
  \caption{Residential Streets in Salt Lake City (Green is residential, Gray is non-residential)}\label{fig:res-nonres}
  \end{subfigure}
  \begin{subfigure}{.45\textwidth}
  \centering
  \includegraphics[width=0.9\textwidth]{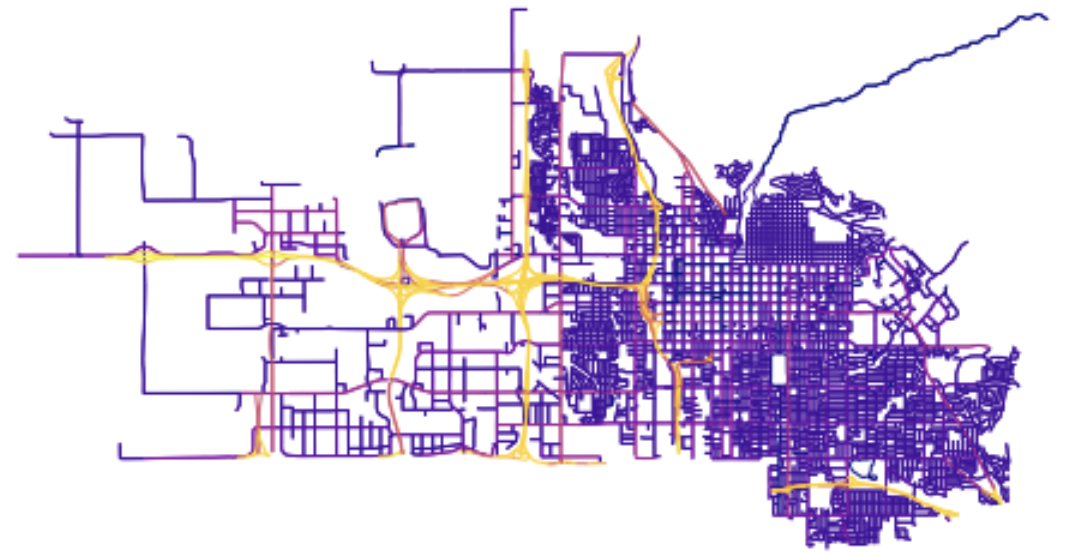}
  \caption{Final Speed Limit Distribution (purple is slow, yellow is fast)}\label{fig:speed-limit-final}
  \end{subfigure}
\end{figure}

Figure \ref{fig:maxspeed} shows a map of the initial \emph{maxspeed} state of Salt Lake City's edges. Some of the edges are labeled in gray, while others are labeled in red. The red edges are initially null, while the gray edges are those with a maxspeed property already available on OSM. Figure \ref{fig:res-nonres} diagrams the residential (green) and non-residential (gray) streets in Salt Lake City. Through these maps, it is clear that imputing residential streets with the default speed limit takes cares of imputing the vast majority of streets within the city. Figure \ref{fig:speed-limit-final} shows what the speed distribution looks like in Salt Lake City after all speed limits are imputed. The coloring is graded from purple, which is a speed limit of 20, to yellow, which is a speed limit of 75+.

The limits are imputed using the above method. There are some limitations to this approach. The first is that the accuracy of speed limits is directly correlated with the accuracy of lane counts and highway types on OSM, which is publicly sourced data that anyone can change. The second limitation lies in our assumption that every edge along a road (given by name) defaults to the same speed limit, which is not necessarily accurate since the 90\% threshold is arbitrary. The third limitation is that we assume that the combination of lane counts and road types are enough to determine speed limit; if there are two speed limits with similar frequencies under a category, for example, we still assume that every null street in that category corresponds with the most frequent speed limit. Lastly, the implicit assumption in logic speed limits is that speed limits are an accurate estimation of the maximum possible speed on a road; this is not necessarily true, especially in cities with several undefined speed limits or those where determined speed limits are inaccurate to the road's actual properties.

\subsection{Gradient Data}
Aside from imputing the values of attributes already available on OpenStreetMap, there are many street-level variables that may also be of interest to transportation researchers. One such attribute is road gradient, which greatly affects driver behavior and isn't directly encoded in an intersection's geometry. 

\begin{figure}[ht]
  \centering
  \includegraphics[width=0.6\textwidth]{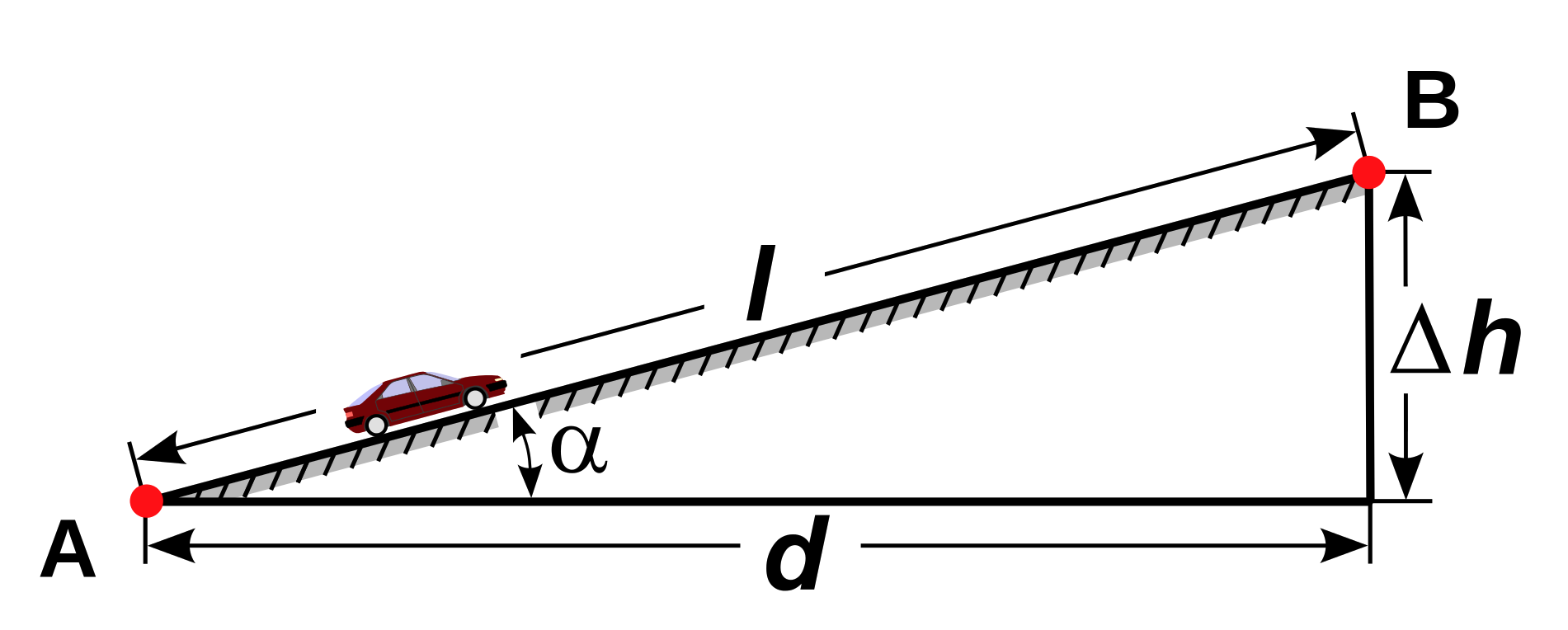}
  \caption{Road Grade Definition}\label{Fig:grade-diagram}
\end{figure}

In Figure \ref{Fig:grade-diagram}, the road's grade (also known as slope, incline or gradient) is defined as $\frac{\Delta h}{d}$~\cite{grade-diagram}. We can call $d$ the road's \emph{run} and call the quantity $\delta h$ the road's \emph{rise}. Because the ratio of rise over run is often quite slight, a road's grade is often expressed as a \emph{percent grade}, defined as $100 \times \frac{\texttt{rise}}{\texttt{run}}$. In practice, our team used the \emph{length} attribute on OSM as the road's \emph{run} and computed the road's $rise$ as the difference between its endpoint's elevations. Each node's elevation is determined by a call to the US Geological Survey's Elevation Point Query Service.~\cite{elevation-point-query-service}.

\begin{figure}[!ht]
  \centering
  \begin{subfigure}{.45\textwidth}
  \centering
  \includegraphics[width=1\textwidth]{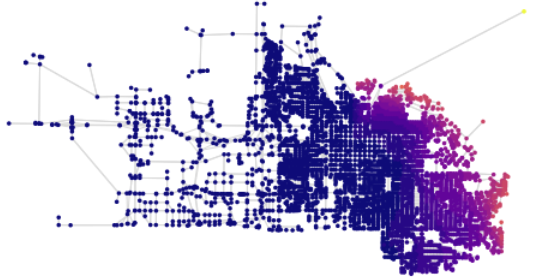}
  \caption{Elevation from USGS}\label{fig:elevation}
  \end{subfigure}
  \begin{subfigure}{.45\textwidth}
  \centering
  \includegraphics[width=1\textwidth]{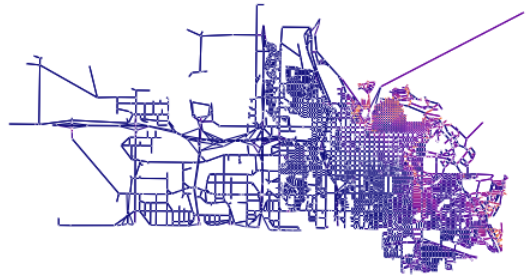}
  \caption{Compute Road Grades (absolute)}\label{fig:grade-map}
  \end{subfigure}
  \caption{Imputed Elevation and Road Grade}
  \label{Fig:grade-elevation}
\end{figure}

Figure \ref{Fig:grade-elevation} shows a distribution of absolute edge grades and node elevations through Salt Lake City. This figure is consistent with the ground truth of Salt Lake City's geography; the city is bordered in the north-east by the Wasatch mountain range, so the increase in road grade and elevation in the north-east corner of the city is expected.~\cite{visit-salt-lake}.

\section{Results} \label{results}
In this section, we initially demonstrate how our pipeline is able to integrate OSM and generate comprehensive representations of signalized intersections. We then evaluate the performance of our model by making comparisons with the ground truth obtained by examining the satellite view from Google earth and Google Street View. In the end, we take a step further and discuss how such datasets can provide insights on downstream applications. 

\begin{figure}[!ht]
  \centering
  \begin{subfigure}{.42\textwidth}
  \centering
  \vspace{6mm}
  \includegraphics[width=1\textwidth]{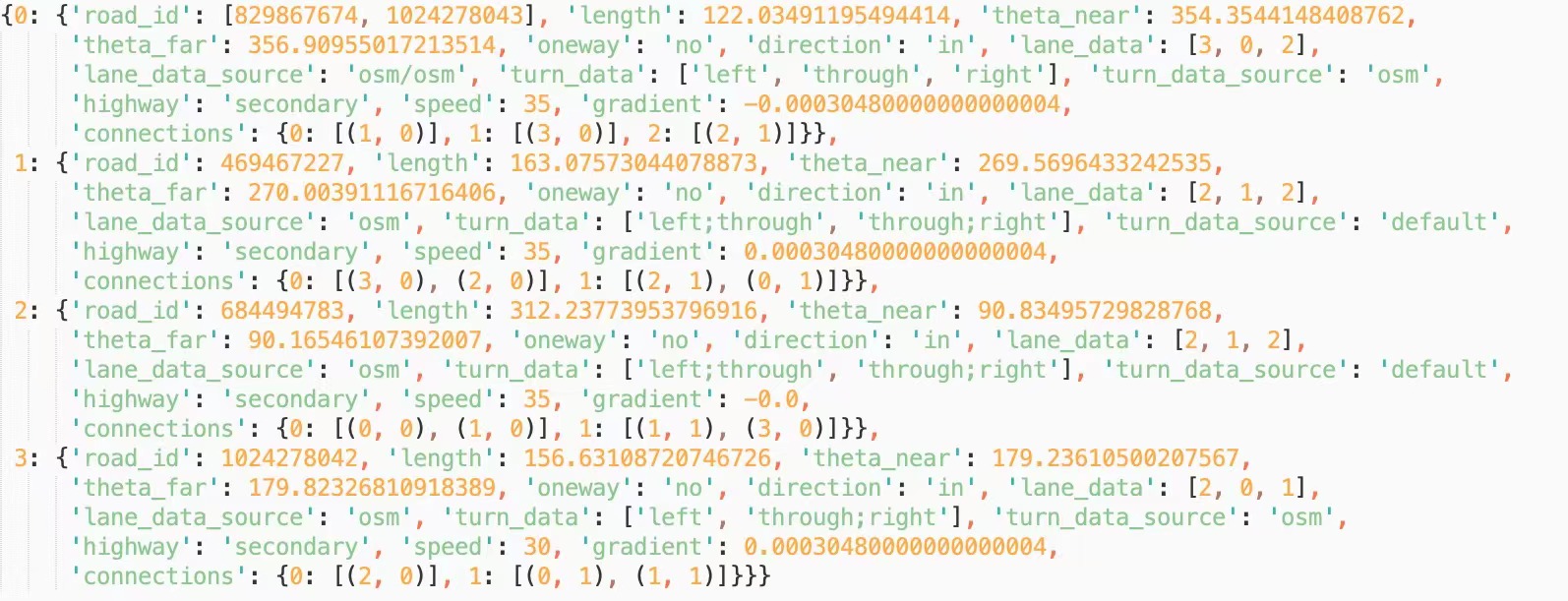}
  \vspace{7mm}
  \caption{An example output for one signalized intersection}\label{Fig:result:example(1)}
  \end{subfigure}
  \begin{subfigure}{.3\textwidth}
  \centering
  \vspace{2mm}
  \includegraphics[width=1\textwidth]{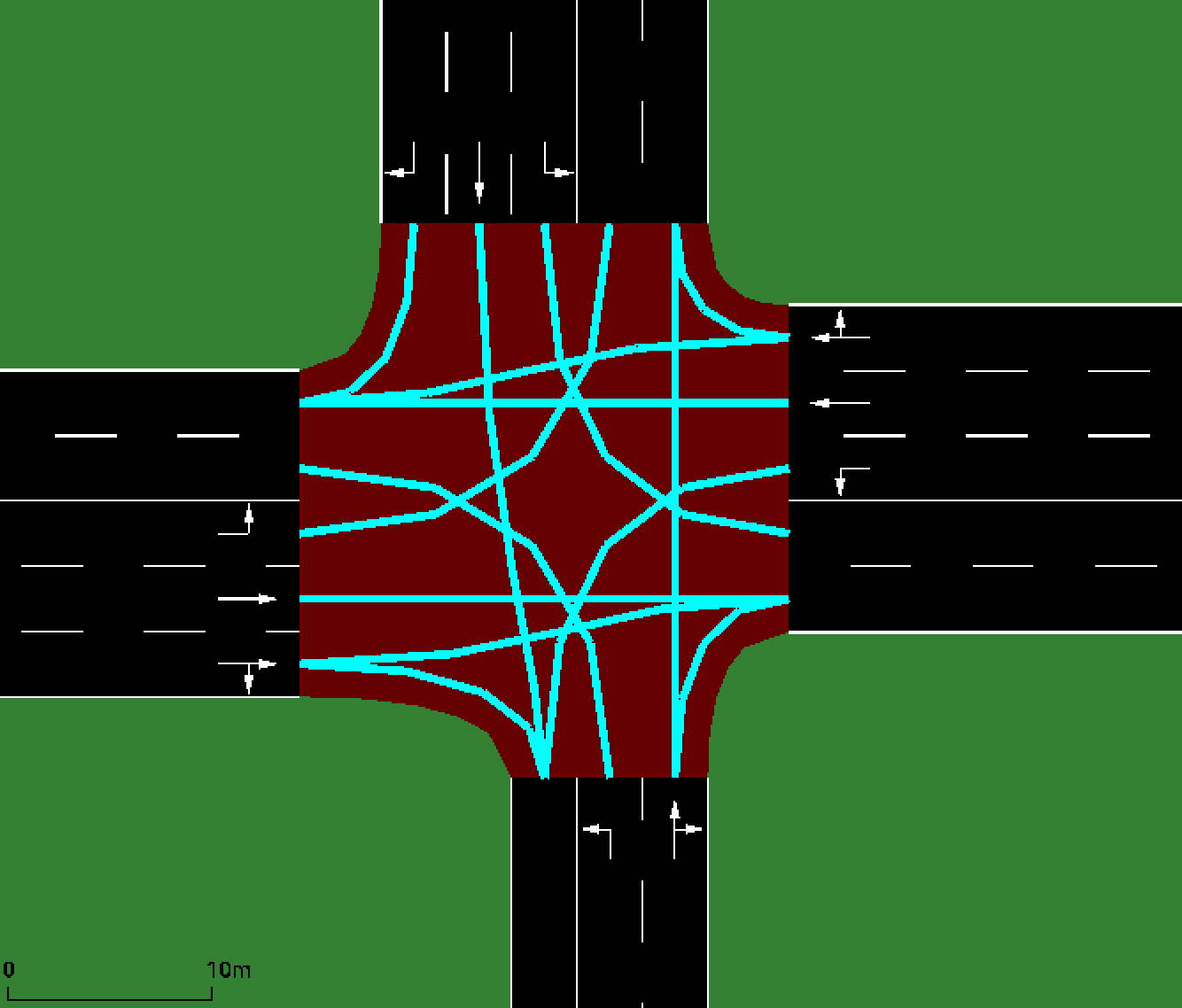}
  \caption{Display of the generated SUMO file}\label{Fig:result:example(2)}
  \end{subfigure}
  \begin{subfigure}{.26\textwidth}
  \centering
  \includegraphics[width=1\textwidth]{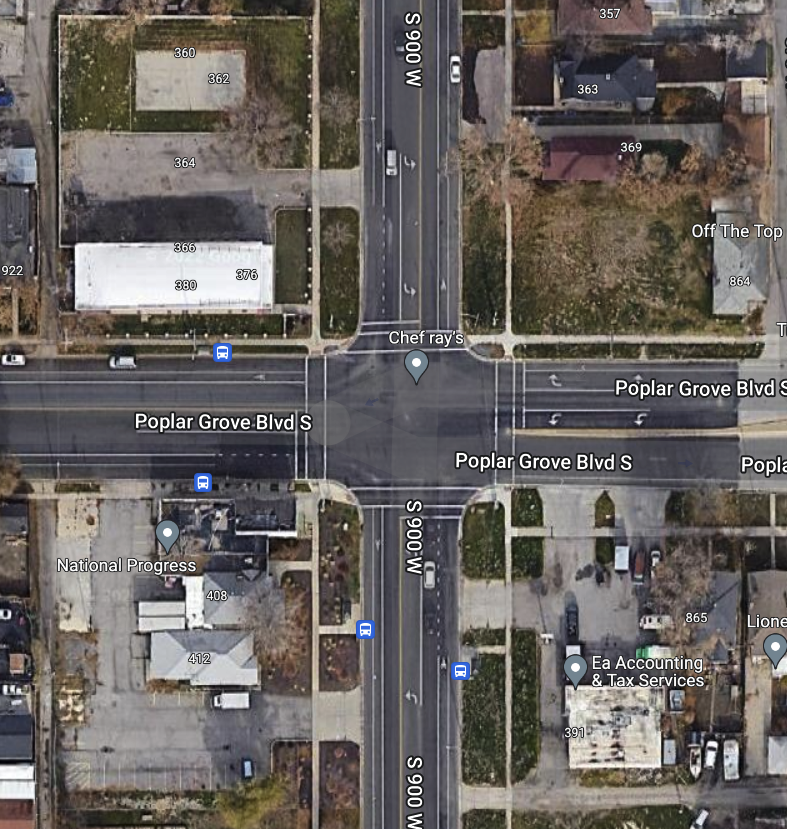}
  \caption{Satellite view of the signalized intersection}\label{Fig:result:example(3)}
  \end{subfigure}
  \caption{Satellite view as a reference}
  \label{Fig:result:example}
\end{figure}

\subsection{An output example}
For demonstration purposes, we randomly picked a signalized intersection of Poplar Grove Boulevard and South 900 West in Salt Lake City. A satellite view of the intersection is shown in Figure \ref{Fig:result:example(3)}. Our pipeline is able to convert the raw data with 43.5\% attributes missing into the formatted JSON file as shown in (a) where all important attributes, either collected from OSM or imputed based on our assumptions, are concisely organized. We also developed a tool to convert such outputs into SUMO files to facilitate usages for various studies, which was included as a function \texttt{gen\_net} of our open source Python library \href{https://pypi.org/project/OSMint/}{OSMint}. The JSON output and converted SUMO file are presented in Figure \ref{Fig:result:example(1)} and \ref{Fig:result:example(2)}, respectively.

\subsection{Performance evaluation}

\subsubsection{Degree distribution}
\begin{figure}[ht!]
  \centering
  \begin{subfigure}{.65\textwidth}
  \centering
  \includegraphics[width=1\textwidth]{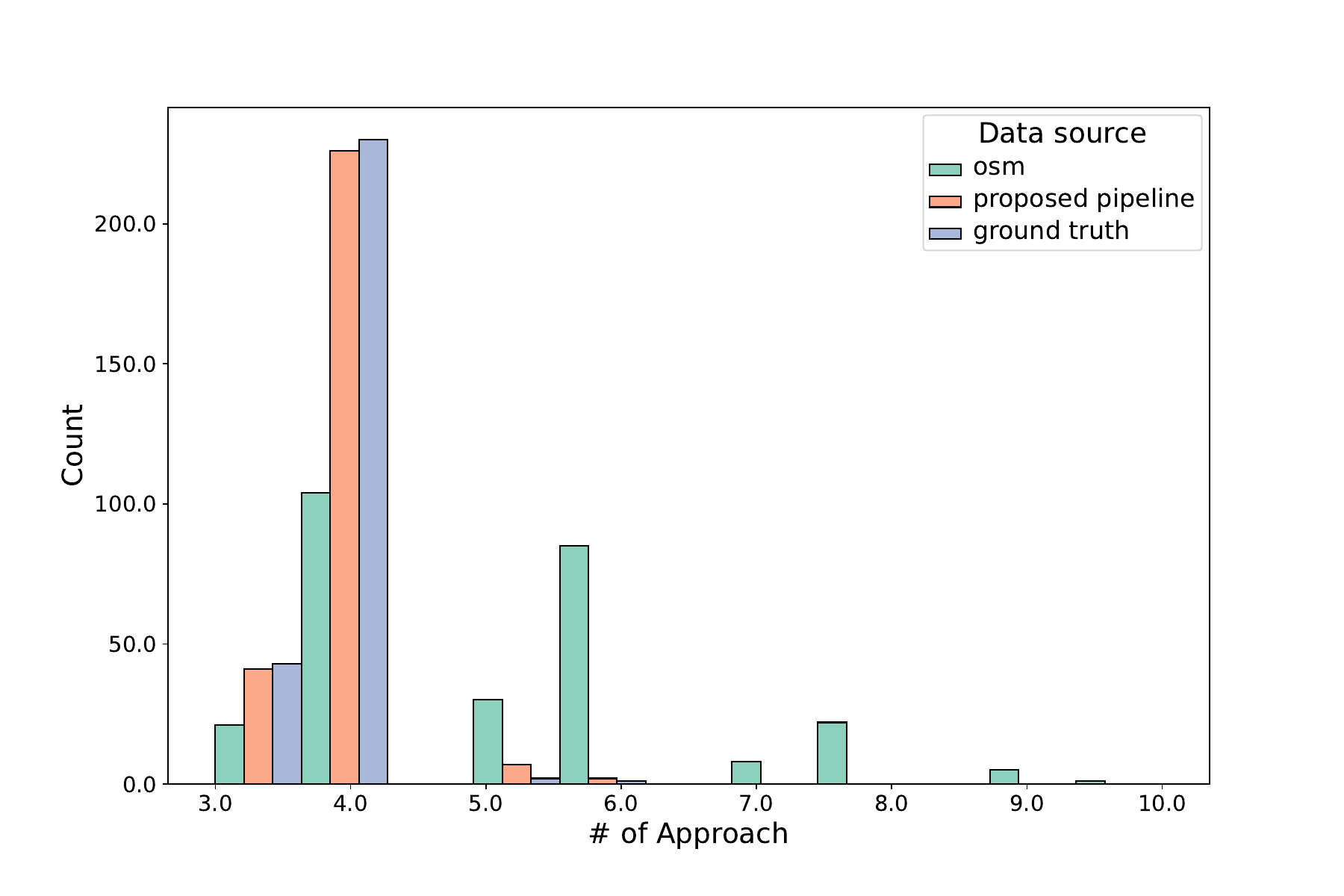}
  \caption{Degree distributions obtained from different data sources}\label{Fig:result:degree(1)}
  \end{subfigure}
  \begin{subfigure}{.4\textwidth}
  \centering
  \includegraphics[width=1\textwidth]{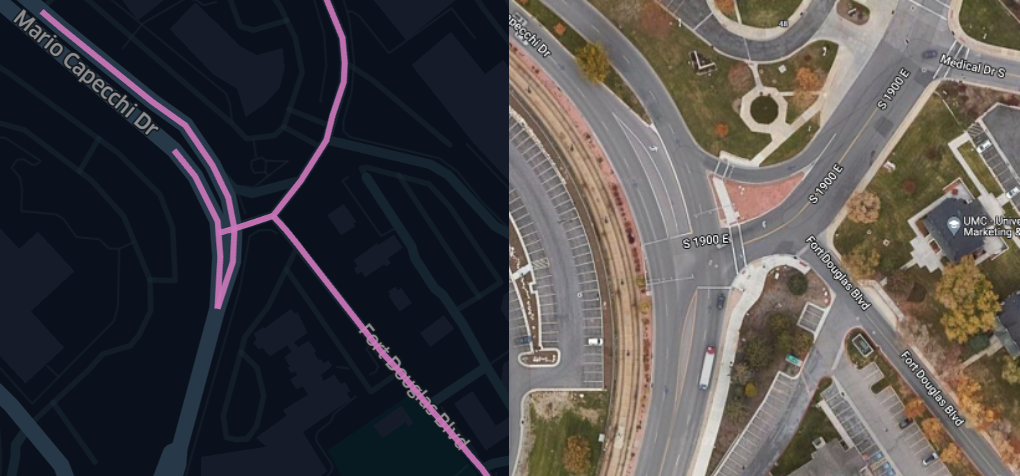}
  \caption{Our method: 4 vs Ground truth: 3}\label{Fig:result:degree(2)}
  \end{subfigure}
  \begin{subfigure}{.4\textwidth}
  \centering
  \includegraphics[width=1\textwidth]{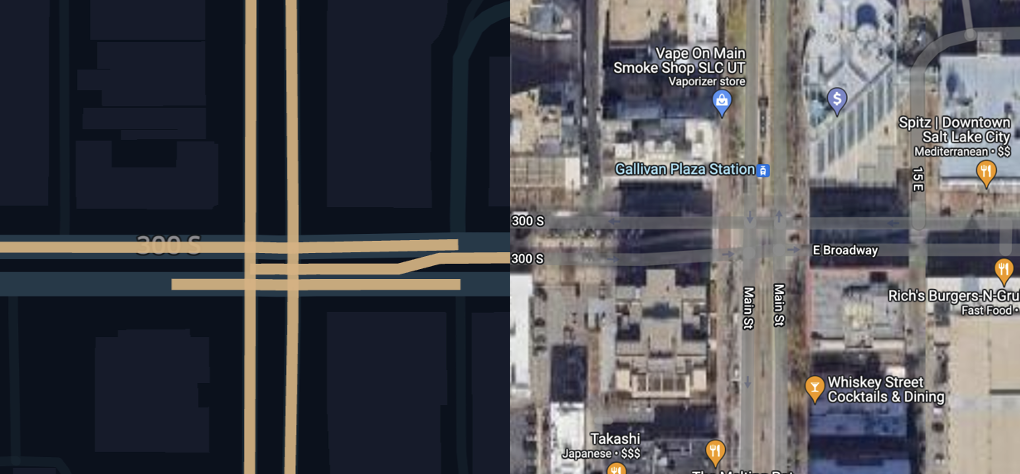}
  \caption{Our method: 5 vs Ground truth: 4}
  \label{Fig:result:degree(3)}
  \end{subfigure}
  \caption{A comparison on degree distributions and analysis on failure cases}
  \label{Fig:result:degree}
\end{figure}

A key functionality of our pipeline is to combine anti-parallel road segments that could have been represented as one two-way. This feature is very helpful for analyzing the distribution of different types of signalized intersections. Without this step, a signalized intersection formed by two divided roads would contain eight approaches and we would not be able to recognize it as a four-way intersection. To evaluate the performance of our method, we compare the obtained distribution of intersection degrees (number of approaches) with the ground truth and the results are shown in Fig \ref{Fig:result:degree(1)}. We can observe that our proposed pipeline obtained a very similar degree distribution to ground truth while the distribution obtained by naively counting the number of connected OSM road segments is very inaccurate. Out of 277 signalized intersections we identified, only 8 intersections failed to match the ground truth degree. We found that those failures mostly fall into the two categories shown in Figure \ref{Fig:result:degree(2)} and \ref{Fig:result:degree(3)}. The first type of failure is because some non-signalized intersections are too close (even closer than a very small $\delta=10m$ we picked) and our pipeline is not able to exclude roads connected to such non-signalized intersections. The second type of failure is more about editing issues on OSM. In the figure, we can see that there is a redundant road segment in OSM that is also connected to the signalized intersection, which leads to a count of 5 approaches instead of 4.

\subsubsection{Intersection attributes}
In addition to the distribution of intersection degrees, we also comprehensively evaluated the performance of our pipeline on other attributes. As shown in Table 2, for 277 signalized intersections with 1381 road segments involved in Salt Lake City, we were able to make a total of 2923 imputations for missing attributes based on assumptions we discussed in Section \ref{methods}. We have also assessed the accuracy of all imputations we made for lane count and turns by comparing with the ground truth. Although the accuracy scores are not very high, they are decent improvements on data quality considering the high missing rates. It is not easy to obtain the ground truth data for the other three attributes we examined and we will leave that as a future work.

\begin{table}[H]
\resizebox{\textwidth}{!}{%
\begin{tabular}{|l|lllll|}
\hline
                   & Lane Data & Geometry Recovered & Turn Data & Speed Limit & Gradient \\ \hline
\# of changes      & 142       & 754                & 1202      & 263         & 1306     \\ \hline
Percentage changed & 10.78\%   & 57.21\%            & 91.20\%   & 19.96\%     & 99.19\%  \\ \hline
Accuracy           & 59.70\%    & N/A                & 58.33\%    & N/A         & N/A      \\ \hline
\end{tabular}%
}
\caption{Performance evaluation of our purposed pipeline}
\label{Table:eval}
\end{table}
\subsection{Analysis on Output Distribution}
\begin{figure}[H]
  \centering
  \includegraphics[width=0.8\textwidth]{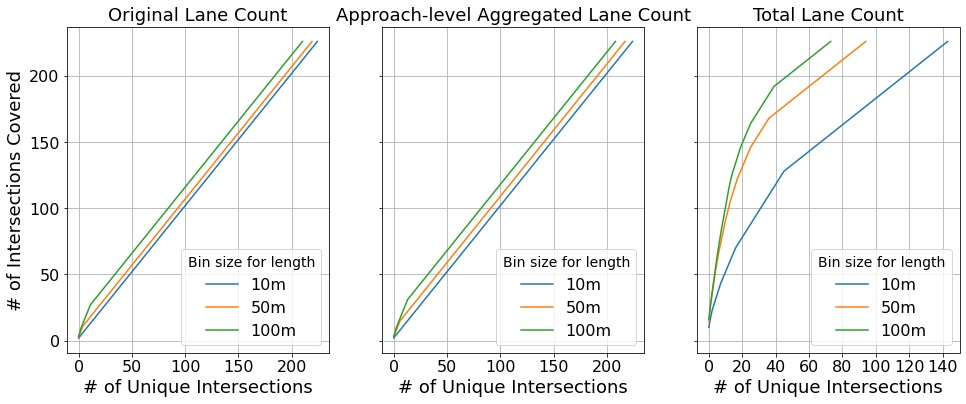}
  \caption{}
  \label{Fig:result:distribution}
\end{figure}

A main purpose to develop this pipeline is to provide a standardized approach for extracting data about signalized intersections from OSM and reduce the bias that may exist in current studies focusing at the level of signalized intersection. For example, an arbitrary intersection may not be able to give an accurate assessment for intersection control models. This is because an arbitrary intersection or a small sample cannot represent the population which should be considered as a joint high-dimensional distribution given the fact that a signalized intersection is associated with a great number of features. Therefore, we take one step further and consider each signalized intersection as a sample from this high-dimensional distribution. We use two attributes--lane count and approach length--as an example to explore this statistical distribution for 226 four-way signalized intersections in Salt Lake City. As shown in Figure \ref{Fig:result:distribution}, for each of two types of attributes, we use three levels of granularity to process them. For approach length, we put the length of each approach into bins with sizes of 10m, 50m, and 100m. For lane count, we consider original lane count (incoming and outgoing for each approach), aggregated count at approach-level, and total lane count at intersection level. The results show that even if we only consider two types of features, the joint distribution is very flat and we need many unique intersections to cover most of the population. In reality, a signalized intersection is defined by a lot more attributes, including some dynamic characteristics such as weather and traffic flow. We hope that this exploration will inspire researchers to work towards generalizability when applying models to signalized intersections, eventually speeding up society's path toward a future with smarter, greener, and safer transportation. 

\section{Limitations and Future Work}
This study has some limitations. First, all of the imputation methods that we have used to create our intersection distribution of Salt Lake City rely directly on our assumption that OpenStreetMap's  data on Salt Lake City's streets have no errors. We make a similar assumption when computing gradient data, as we assume that the US Geological Survey's elevation data is fully accurate to the ground truth, as well as that every edge can be modeled as a straight line (to use length in our gradient calculation). Second, our imputation methods for lane data, speed limits, geometry, and especially turn data impute large proportions of the data based on a small minority of intersections that have data availability. This assumes that the available data can be used to accurately model the remaining intersections for which data is not available, without taking other potential factors into consideration. 

\label{future_work}
One potential direction to explore in future work is analyzing this paper's proposed methods with respect to other cities, both within the United States and around the world. Salt Lake City is a relatively small and low-density city compared to more populated American metropolises like New York and Los Angeles, so investigating the effectiveness of this paper's methods on other cities may provide some insight into the impact of other factors, such as the prevalence of bicyclists in European cities or the dominance of motorcycles on Indian roads.

Another important place to expand upon the work communicated in this paper is in devising more accurate methods to impute lane count and turn data. Since OSM's data may be fundamentally inaccurate, using a computer vision algorithm that incorporates both satellite imagery and StreetView to impute lane counts and turning data, for example, would be a very accurate way to collect ground truth data at each intersection without relying on the unreliability that comes from crowd-sourcing. Incorporating street-level camera footage would help mitigate the concerns of obscured satellite images due to buildings, trees, or other obstacles. 

\section{Conclusion} \label{conclusion}
In conclusion, signalized intersections are crucial to making intelligent transportation systems safer, cleaner, and more reliable. Inspired by the impact that OpenStreetMap has had in enabling transportation research around the world, we hope that our proposed method to extract traffic intersections from raw crowd-sourced data will provide researchers with the tools to create real-life intersection distributions for their own innovative applications. We also hope that more efforts will be put into building standardized and systematic data pipelines for different transportation research problems. \\
Our package can be found at the following link: \url{https://pypi.org/project/OSMint/}

\newpage
\bibliographystyle{trb}
\bibliography{trb_template}

\end{document}